\documentclass[twocolumn,aps,prc,superscriptaddress,showpacs,floatfix]{revtex4}

\usepackage{epsfig}
\usepackage{amsmath}
\usepackage{multirow}
\usepackage{graphicx}
\usepackage{amsmath}
\usepackage{feynmf}
\usepackage[normalem]{ulem}  
\usepackage[dvips]{color} 

\renewcommand\sout{\bgroup \color{red} \ULdepth=-.5ex \ULset}

\newcommand{\Ex}[2]{\ifmmode{#1\times10^{#2}}\else{$#1\times10^{#2}$}\fi}

\begin{document}
\title{Color spin wave functions of heavy tetraquark states}

\author{Woosung Park}\affiliation{Department of Physics and Institute of Physics and Applied Physics, Yonsei
University, Seoul 120-749, Korea}
\author{Su Houng~Lee}\affiliation{Department of Physics and Institute of Physics and Applied Physics, Yonsei
University, Seoul 120-749, Korea}
\date{\today}
\begin{abstract}
Using the variational method, we calculate the mass of the $J^P=1^+$  $ud\bar{b}\bar{b}$ tetraquark containing two identical heavy antiquarks in a nonrelativistic potential model with color confinement and spin hyperfine interaction.
In particular, we extend a previous investigation of the model by Brink and Stancu by investigating the effect of including the color anti-sextet component of the  diquark configuration as well as using several more Gaussian parametrization for the  L=0 part of the spatial wave function.
We find that for the heavy tetraquark, the $6\bar{6}$ component among  the color singlet bases is negligible and that the previously used specific Gaussian spatial configuration is good enough in obtaining the ground state energy.
\end{abstract}


\maketitle

\section{Introduction}

Recently, several new heavy mesons were discovered  with masses  difficult to explain within the conventional quark model and thus could either be a  multiquark or a molecular configuration~\cite{Nielsen:2009uh}.  These are the $D_s$ states, X(3872)~\cite{Choi:2003ue}, Z(4051) and Z(4248)~\cite{Mizuk:2008me}, and the newly discovered charged charmonium like states $Z_c$(3900)\cite{Ablikim:2013mio,Liu:2013dau,Xiao:2013iha}. While the recently observed charged states are mostly likely of  exotic configurations, their quantum numbers are not explicitly exotic. On the other hand, there are a number of works suggesting that  certain flavor exotic multiquark states with heavy quarks could be stable under strong decay and be observable from B-decay or heavy ion collisions. If such particles are indeed found, they would mark the first observation of flavor exotic multiquark configuration, which will lead to a new dimension of hadron spectroscopy\cite{Lee:2007tn,Lee:2009rt}.

The first set of papers suggesting the  tetraquark configurations were given by Jaffe~\cite{Jaffe:1976ih,Jaffe:1976ig}, within the MIT bag model with color spin interaction.  This paper subsequently promoted an intense  discussion on the
possible existence of tetraquark states. It was suggested in his papers that the   $f_0(975)$ and $a_0(980)$ resonances could be interpreted as part of the scalar $J^{PC}=0^{++}$ nonet  composed of $qq\bar{q}\bar{q}$ tetraquarks.  This picture was later further confirmed by Weinstein and Isgur~\cite{Weinstein:1983gd,Weinstein:1990gu}, establishing the possible existence of tetraquark in a variety of quark models. This means that tetraquarks with heavy quarks can also exist.
In fact, the calculation for the spectrum of $c\bar{c}q\bar{q}$ tetraquark which was performed by Stancu~\cite{Stancu:2006st} and Hogassen ~\cite{Hogaasen:2005jv} suggest that X(3872) meson which have been discovered by Belle~\cite{Choi:2003ue} could be a  $c\bar{c}q\bar{q}$ tetraquark state.  This state however is of the cryptoexotic nature, with hidden heavy flavor quantum number.  Moreover, these states could be a meson-meson bound molecular states as was predicted more than twenty years ago\cite{Tornqvist:1993ng}. Thus it is experimentally a challenge to prove that they are composed of purely tetraquark components.  

Simple estimates based on color-spin interaction suggests that there could be stable heavy tetraquark states with explicitly flavor exotic quantum number\cite{Lee:2007tn,Lee:2009rt}.  In particular, the $J^P=1^+,I=0$ $ud\bar{Q}\bar{Q}$, with $Q$ being a heavy quark and called the $T_{QQ}^1$, are of particular interest as it could be a stable flavor exotic tetraquark\cite{Zouzou:1986qh} state that could be produced in electro-positron collision\cite{Hyodo:2012pm} or in a heavy ion collision\cite{Cho:2010db}.  The stability of $T_{QQ}^1$ has been studied in quark model~\cite{Cohen:2006jg,Vijande:2007ix,Vijande:2007rf,Lee:2009rt} and QCD
sum rules~\cite{Navarra:2007yw}.            .

Here, we are interested in elaborating the quark model calculation for $T_{QQ}^1$,  obtained  with the  nonrelativistic potential as given by Silvestre-Brac and Semay ~\cite{SilvestreBrac,SilvestreBrac:1993ry}, that was performed by Brink and Stancu (BS) ~\cite{Brink:1998as} using  the variational method based on simple Gaussian trial function which is useful to describe nuclear few-body systems~\cite{Kamimura:1988zz}.
The mass of $T_{bb}^1$ calculated by BS was  33$ ~{\rm MeV}$ above the results by Silvestre-Brac and Semay~\cite{SilvestreBrac,SilvestreBrac:1993ry} that used a variational calculation with many oscillator bases.
BS proposed several alternatives of improving the variational energy in their calculation.
In this work, we extend the work of BS by investigating their proposal of improvements.  In the first improvement, since Brink and Stancu~\cite{Brink:1998as} excluded the $6\bar{6}$ component in color singlet basis following the assumption given in~\cite{Zouzou:1986qh}, we explicitly investigate the validity of the assumption by including the $6\bar{6}$ component in the calculation. This calculation will be performed with the same single Gaussian spatial wave function as was done by BS that will  be called scheme 0 in our work.  In the second improvement, we extend the simple spatial configuration used by BS  to the generalized cases introduced as scheme I to V in section IV. This is to investigate the extended correlations between quarks.  We further introduce schemes(scheme VI and VII) to investigate the importance of using multiple Gaussian to the wave function. The simple Gaussian function for total angular momentum L=0  is convenient to examine the variational energy of the tetraquark containing two identical heavy antiquarks in such a situation. We found that the size of tetraquark is important to understand the stability of heavy tetraquark.
We also calculate the quark-antiquark meson masses within the same model parameters. Using these results, we investigate the stability of the tetraquark states against the decay into two meson states.

In section II we introduce the hamiltonian.  In section III, we introduce the  spatial and color-spin wave function.  In section IV, we introduce the different schemes and calculate the matrix elements.  In section V, we show the numerical results and discuss the two improvements.  In section VI, we analyze the mass splitting coming from hyperfine potential.  Finally, we give the summary in section VII.

\section{Hamiltonian}

Let us start from a nonrelativistic Hamiltonian, that includes  confinement and hyperfine potential for the color and spin degrees of freedom:
\begin{eqnarray}
H=\sum_{i=1}^{4}(m_{i}+\frac{\textbf{p}^2_i}{2m_i})-\frac{3}{4}\sum_{i<j}^{4}
\frac{\lambda^c_i}{2}\frac{\lambda^c_j}{2}(V^{C}_{ij}+V^{SS}_{ij}). \label{Hamiltonian}
\end{eqnarray}
Here, $m_i$'s are the quark masses; $\lambda^c_i/2$ are the color operator of the $i$'th quark for the color SU(3); $V^{C}_{ij}$ and $V^{SS}_{ij}$ are the confinement and hyperfine potential, respectively.   We adopt the confinement and hyperfine potential from ref.~\cite{Bhaduri:1981pn}:
\begin{eqnarray}
V^{C}_{ij}=-\frac{\kappa}{r_{ij}}+\frac{r_{ij}}{a_0^2}-D, \label{vc_ij}
\end{eqnarray}
\begin{eqnarray}
V^{SS}_{ij}=\frac{\hbar^2c^2\kappa}{m_im_jc^4}\frac{1}{r_0^2r_{ij}}e^{-r_{ij}/r_0}
{\sigma}_i\cdot{\sigma}_j,
\end{eqnarray}
where $r_{ij}=\mid\textbf{r}_i-\textbf{r}_j\mid$ and ${\sigma}_i$ is the spin operator.

Since our aim is to generalize and compare with the calculation of BS~\cite{Brink:1998as} for the mass of the tetraquark containing two light quarks and two heavy antiquarks with variational method, for the  parameters appearing in Eqs.(2)-(3), we chose the same values as those used in that paper.
The parameters are given by
\begin{align}
&m_u=m_d=337  ~{\rm MeV}, \nonumber \\
&m_c=1870 ~{\rm MeV},~~~ m_b=5259 ~{\rm MeV},\nonumber
\end{align}
\begin{eqnarray}
\kappa=102.67 ~{\rm MeV}~{\rm fm},\qquad a_0=0.0326 \quad (~{\rm MeV^{-1}}~{\rm fm})^{1/2},\nonumber
\end{eqnarray}
\begin{eqnarray}
D=913.5 ~{\rm MeV},\qquad r_0=0.4545 ~{\rm fm}.
\end{eqnarray}

\section{Wave function}

In this work, we will be interested in the $T_{QQ}^1$ state within the Hamiltonian introduced above.  In the constituent quark model, the lowest mass for the $T_{QQ}^1$ state is obtained in a configuration where all the quarks are in the $l=0$ state.
Therefore, the Hamiltonian introduced in the previous section will be applied to only the $s$-wave configurations that depend also on  the color and spin states. Now, we establish the appropriate
basis functions for describing the  tetraquark system.

\subsection{spatial function}
In order to use variational method, we construct the trial wave function for the spatial part in a  simple
Gaussian form. This spatial function makes it easy to calculate the matrix element of the Hamiltonian. When we calculate the matrix element of the potential terms for the tetraquark configuration with certain symmetry,  it is convenient to introduce the following three coordinate configurations which are related with each other by orthogonal matrix.
\begin{itemize}
\item Coordinate I :
\begin{eqnarray}
\pmb{ \rho}=\frac{1}{\sqrt{2}}(\textbf{r}_1-\textbf{r}_3),\qquad
\pmb{\rho}^{\prime}=\frac{1}{\sqrt{2}}(\textbf{r}_2-\textbf{r}_4),\nonumber
\end{eqnarray}
\begin{eqnarray}
\pmb{x}=\frac{1}{2}(\textbf{r}_1-\textbf{r}_2+\textbf{r}_3-\textbf{r}_4).
\end{eqnarray}
\item Coordinate II :
\begin{eqnarray}
\pmb{ \alpha}=\frac{1}{\sqrt{2}}(\textbf{r}_1-\textbf{r}_4),\qquad
\pmb{ \alpha}^{\prime}=\frac{1}{\sqrt{2}}(\textbf{r}_2-\textbf{r}_3),\nonumber
\end{eqnarray}
\begin{eqnarray}
\pmb{y}=\frac{1}{2}(\textbf{r}_1-\textbf{r}_2-\textbf{r}_3+\textbf{r}_4).
\end{eqnarray}
\item Coordinate III :
\begin{eqnarray}
\pmb{ \sigma}=\frac{1}{\sqrt{2}}(\textbf{r}_1-\textbf{r}_2),\qquad
\pmb{\sigma}^{\prime}=\frac{1}{\sqrt{2}}(\textbf{r}_3-\textbf{r}_4),\nonumber
\end{eqnarray}
\begin{eqnarray}
\pmb{\lambda}=\frac{1}{2}(\textbf{r}_1+\textbf{r}_2-\textbf{r}_3-\textbf{r}_4).
\end{eqnarray}
\end{itemize}
Here, particles 1 and 2 indicate quarks, while 3 and 4 indicate antiquarks.

When describing the diquark-antidiquark system, it is convenient to choose the closed form coordinate III.
Hence, for calculating the matrix element of the Hamiltonian, we use coordinate III. On the other hand, it is convenient to choose coordinates I or II in describing the asymptotic form corresponding  to either the direct or exchange meson-meson system. As we deal with the tetraquark consisting of two identical antidiquark, we must consider the permutation of (12) and (34) with respect to the basis function. In other words, we must construct the bases functions satisfying the Pauli principle. For these three coordinate configurations under the permutation of (12) and (34), we obtain the following property:
\begin{eqnarray}
(12)\pmb{ \rho}=\pmb{ \alpha}^{\prime},\quad
(12)\pmb{\rho}^{\prime}=\pmb{ \alpha},\quad
(12)\pmb{\sigma}=-\pmb{\sigma},\quad(12)\pmb{\lambda}=\pmb{\lambda},\nonumber
\end{eqnarray}
\begin{eqnarray}
(34)\pmb{ \rho}=\pmb{ \alpha},\quad
(34)\pmb{\rho}^{\prime}=\pmb{ \alpha}^{\prime},\quad
(34)\pmb{{\sigma}^{\prime}}=-\pmb{{\sigma}^{\prime}},\quad
(34)\pmb{\lambda}=\pmb{\lambda}. \nonumber \\
\end{eqnarray}

We denote the spatial function by $R^{s}$ which has been introduced by BS in Ref~\cite{Brink:1998as}. As was discussed by BS, the most general Gaussian form for the L=0 spatial function can be written in terms of six scalar quantities as given by
\begin{eqnarray}
R^{s}=\exp[-(C^s_{11}\sigma^2+C^s_{22}\sigma^{\prime2}+C^s_{33}\lambda^2+
2C^s_{12}\pmb{\sigma}\cdot\pmb{\sigma^{\prime}} \nonumber \\
+2C^s_{13}\pmb{\sigma}\cdot\pmb{\lambda}+
2C^s_{23}\pmb{\sigma^{\prime}}\cdot\pmb{\lambda})]. \label{space-wave-function}
\end{eqnarray}
In order to calculate the  matrix element of the confinement and hyperfine potential terms involving $r_{ij}$, where $i,j=1 \sim4$, it is convenient to
represent the argument of the exponential function in a matrix form so that one can easily transform from one coordinate to the other by orthogonal transformations.
Therefore, we define the coordinate configurations in a matrix form as follows:
\begin{equation}
X=\left(\begin{array}{c}\pmb{\rho} \\  \pmb{\rho^{\prime}} \\ \pmb{x}\end{array} \right),\quad
Y=\left(\begin{array}{c}\pmb{\alpha} \\  \pmb{\alpha^{\prime}} \\ \pmb{y}\end{array} \right),
\quad
Z=\left(\begin{array}{c}\pmb{\sigma} \\  \pmb{\sigma^{\prime}} \\ \pmb{\lambda}\end{array} \right).
\end{equation}
Then, we can write $R^s$ of Eq.~(9) in the following form
\begin{eqnarray}
R^{s}=\exp(-Z^TC^sZ),
\end{eqnarray}
where $C^s$ is the symmetric matrix, and $Z^T$ is the transpose of the column matrix $Z$.
Using the orthogonal matrices which transform one coordinate into the other, the $R^s$ can be
expressed in terms of the coordinates (5) and (6). It becomes
\begin{align}
R^{s}&=\exp(-Z^TC^sZ)=\exp(-X^TA^sX)\\  \nonumber
&=\exp(-Y^TB^sY),
\end{align}
where the symmetric matrices $A^s$ and $B^s$ are obtained from the similarity transformation.
Applying the orthogonal matrices to the coordinates and $C^s$ matrix give
\begin{eqnarray}
X=U_xZ,\qquad \quad Y=U_yZ,\\ \nonumber
A^s=U_xC^sU_x^-1, \qquad B^s=U_yC^sU_y^-1,
\end{eqnarray}
where the orthogonal matrices $U_x$ and $U_y$ are
\begin{equation}
U_x=\left(\begin{array}{ccc}\frac{1}{2} & -\frac{1}{2}
&\frac{1}{\sqrt{2}}\\
-\frac{1}{2} & \frac{1}{2} & \frac{1}{\sqrt{2}}\\
\frac{1}{\sqrt{2}} & \frac{1}{\sqrt{2}} & \scriptstyle{0} \end{array} \right),\nonumber
\end{equation}
\begin{equation}
U_y=\left(\begin{array}{ccc}\frac{1}{2} & \frac{1}{2}
&\frac{1}{\sqrt{2}}\\
-\frac{1}{2} & -\frac{1}{2} & \frac{1}{\sqrt{2}}\\
\frac{1}{\sqrt{2}} & -\frac{1}{\sqrt{2}} & \scriptstyle{0} \end{array} \right).
\end{equation}

Introducing the position vector of the center of mass, $\textbf{r}_C=(1/M)\sum m_i\textbf{r}_i$,
where $M=\sum m_i$, the kinetic part of Eq. (1) can be expressed in the center of mass frame.
We can obtain the kinetic part in the center of mass frame by excluding the kinetic energy of the position vector of the center of mass. The kinetic part in the center of mass frame denoted by $T_c$ can be expressed in terms of coordinate III as follows:
\begin{equation}
T_c=\sum_{i=1}^{4}\frac{\textbf{p}^2_i}{2m_i}-\frac{\textbf{p}^2_{{r}_C}}{2M}
=\frac{\textbf{p}^2_{\sigma}}{2m_1}+\frac{\textbf{p}^2_{{\sigma}^{\prime}}}{2m_3}+
\frac{\textbf{p}^2_{\lambda}}{2m^{\prime}}, \label{KE}
\end{equation}
where $m_1=m_2=m_q$, $m_3=m_4=m_Q$, and $m^{\prime}$ is the reduced mass, $2m_1m_3/(m_1+m_3)$.

\subsection{Spin-color state}

The color space  acting on the $\lambda^c_i\lambda^c_j$ in a given flavor configuration
of the tetraquark can be decomposed according to the irreducible representation of color $SU(3)_c$ as
\begin{equation}
3_c\times3_c\times\bar{3}_c\times\bar{3}_c=\bar{3}_c\times3_c+6_c\times\bar{6}_c+
\bar{3}_c\times\bar{6}_c+6_c\times3_c.
\end{equation}
Color singlet states can be obtained from the first and the second term
in the right hand side of Eq.(16). It is convenient to use following notions introduced in Ref.~\cite{Buccella:2006fn} to denote the two color singlets.
\begin{equation}
(q_1q_2)^{\bar{3}}\otimes(\bar{q}_3\bar{q}_4)^3,\qquad
(q_1q_2)^{6}\otimes(\bar{q}_3\bar{q}_4)^{\bar{6}}
\end{equation}
It follows from the property of irreducible representation of color $SU(3)_c$ that $(q_1q_2)^{\bar{3}}\otimes(\bar{q}_3\bar{q}_4)^3$ is antisymmetric under transposition of $q_1$ and $q_2$ or $\bar{q}_3$ and  $\bar{q}_4$, and $(q_1q_2)^{6}\otimes(\bar{q}_3\bar{q}_4)^{\bar{6}}$ is symmetric under transposition of $q_1$ and $q_2$ or $\bar{q}_3$ and  $\bar{q}_4$. Using the tensor notation~\cite{Stancu:1991rc}, the two color singlets can be written as
\begin{align}
&(q_1q_2)^{\bar{3}}\otimes(\bar{q}_3\bar{q}_4)^3=\frac{1}{\sqrt{12}}
\epsilon^{\alpha\beta\gamma}\epsilon_{\alpha\lambda\sigma}
q_{\beta}(1)q_{\gamma}(2)\bar{q}^{\lambda}(3)\bar{q}^{\sigma}(4),\nonumber\\
&(q_1q_2)^{6}\otimes(\bar{q}_3\bar{q}_4)^{\bar{6}}
=\frac{1}{\sqrt{6}}d^{\alpha\beta\gamma}d_{\alpha\lambda\sigma}
q_{\beta}(1)q_{\gamma}(2)\bar{q}^{\lambda}(3)\bar{q}^{\sigma}(4),
\end{align}
where $d^{\alpha\beta\gamma}$ and $d_{\alpha\beta\gamma}$ are
\begin{align}
&d^{111}=d_{111}=d^{222}=d_{222}=d^{333}=d_{333}=1,\nonumber\\
&d^{412}=d_{412}=d^{421}=d_{421}=d^{523}=d_{523}=d^{532}=d_{532}=\nonumber\\
&d^{613}=d_{613}=d^{631}=d_{631}=\frac{1}{\sqrt{2}}.
\end{align}
These two color singlet states are orthonormal by means of the irreducible representation of color $SU(3)_c$. The orthogonality  can also be simply shown by the vanishing of the multiplication of the anti-symmetric to symmetric color indices. The coefficients can be deduced from Young operators associated with sextet and antisextet which are useful to generate the basis state in a Young diagram.
The two color singlet states can be recombined into another two color singlets   constructed from two quark antiquark pair  of color singlet-singlet and an octet-octet states that are  appropriate for studying the decay properties.

Due to the fact that the irreducible representation of  $SU(2)_s$ for an antiquark with spin=1/2 is equivalent to that of a quark, the spin space of the tetraquark can be represented as $V_{1/2}\times{V_{1/2}}\times{V_{1/2}}\times{V_{1/2}}$, and decomposed into the direct sum of the following parts:
\begin{equation}
V_{1/2}\times{V_{1/2}}\times{V_{1/2}}\times{V_{1/2}}=V_0\times{V_0}+{V_0}\times{V_1}
+{V_1}\times{V_0}+{V_1}\times{V_1},
\end{equation}
where the subscripts indicate the spins. Accordingly, the total spin of the tetraquark  can be S=0, 1 or 2. 

For S=0, there are two independent basis states obtained from $V_0\times{V_0}$ and ${V_1}\times{V_1}$ parts. The corresponding bases are denoted by
\begin{equation}
(\chi_{12})_{s=0}\otimes(\chi_{34})_{s=0},
\qquad(\chi_{12})_{s=1}\otimes(\chi_{34})_{s=1}, \label{basis-S0}
\end{equation}
where particles 1 and 2 imply quarks, and particles 3 and 4 antiquarks. 

For S=1, there are three independent basis states coming from $V_0\times{V_1}$, $V_1\times{V_0}$, and $V_1\times{V_1}$ part. These states are given by
\begin{eqnarray}
(\chi_{12})_{s=0}\otimes(\chi_{34})_{s=1},(\chi_{12})_{s=1}\otimes(\chi_{34})_{s=0}, \nonumber \\
(\chi_{12})_{s=1}\otimes(\chi_{34})_{s=1}.\label{basis-S1}
\end{eqnarray}

For  S=2, there exist only one state coming from $V_1\times{V_1}$ part denoted as
\begin{equation}
(\chi_{12})_{s=1}\otimes(\chi_{34})_{s=1}
\end{equation}

The spin states for S=0 and S=1 are orthonormal, as in the color states. It is important to see the permutation property of the spin states under transposition (12) or (34) because
the wave function has to have a definite symmetry under exchange of identical particles; (34) are identical while (12) becomes identical when extended to the flavor space. Applying the transposition (12) or (34) to the spin states give
\begin{align}
&(12)(\chi_{12})_{s=0}=-(\chi_{12})_{s=0},\qquad(12)(\chi_{12})_{s=1}=(\chi_{12})_{s=1}\nonumber\\
&(34)(\chi_{12})_{s=0}=-(\chi_{12})_{s=0},\qquad(34)(\chi_{12})_{s=1}=(\chi_{12})_{s=1}.
\end{align}

In general, when the symmetry constraint is not imposed, there is a four-dimensional color-spin orthogonal basis for S=0 spanned by the following states:
\begin{align}
\phi_1&=(q_1q_2)^{6}\otimes(\bar{q}_3\bar{q}_4)^{\bar{6}}(\chi_{12})_{s=1}\otimes(\chi_{34})_{s=1}\nonumber\\
&\equiv(q_1q_2)^{6}_{1}\otimes(\bar{q}_3\bar{q}_4)^{\bar{6}}_{1}
,\nonumber
\end{align}
\begin{align}
\phi_2&=(q_1q_2)^{\bar{3}}\otimes(\bar{q}_3\bar{q}_4)^{3}(\chi_{12})_{s=0}\otimes(\chi_{34})_{s=0}\nonumber\\ &\equiv(q_1q_2)^{\bar{3}}_{0}\otimes(\bar{q}_3\bar{q}_4)^{3}_{0},\nonumber
\end{align}
\begin{align}
\phi_3&=(q_1q_2)^{6}\otimes(\bar{q}_3\bar{q}_4)^{\bar{6}}(\chi_{12})_{s=0}\otimes(\chi_{34})_{s=0}\nonumber\\
&\equiv(q_1q_2)^{6}_{0}\otimes(\bar{q}_3\bar{q}_4)^{\bar{6}}_{0}
,\nonumber
\end{align}
\begin{align}
\phi_4&=(q_1q_2)^{\bar{3}}\otimes(\bar{q}_3\bar{q}_4)^{3}(\chi_{12})_{s=1}\otimes(\chi_{34})_{s=1}\nonumber\\ &\equiv(q_1q_2)^{\bar{3}}_{1}\otimes(\bar{q}_3\bar{q}_4)^{3}_{1}. \label{spin0}
\end{align}

Similarly, we use the following six-dimensional color-spin basis for S=1 state: 
\begin{equation}
\psi_1=(q_1q_2)^{6}_{1}\otimes(\bar{q}_3\bar{q}_4)^{\bar{6}}_{1}, \qquad
\psi_2=(q_1q_2)^{\bar{3}}_{1}\otimes(\bar{q}_3\bar{q}_4)^{3}_{1},\nonumber
\end{equation}
\begin{equation}
\psi_3=(q_1q_2)^{\bar{3}}_{0}\otimes(\bar{q}_3\bar{q}_4)^{3}_{1},\qquad
\psi_4=(q_1q_2)^{6}_{1}\otimes(\bar{q}_3\bar{q}_4)^{\bar{6}}_{0},\nonumber
\end{equation}
\begin{equation}
\psi_5=(q_1q_2)^{\bar{3}}_{1}\otimes(\bar{q}_3\bar{q}_4)^{3}_{0},\qquad
\psi_6=(q_1q_2)^{6}_{0}\otimes(\bar{q}_3\bar{q}_4)^{\bar{6}}_{1}. \label{spin1}
\end{equation}

Depending on the tetraquark state, the actual states contributing to the bases  will
be smaller due to symmetry considerations.
Our main interest is in the tetraquark $T^1_{QQ}$ containing two identical heavy antiquarks and two light quarks $u$ and $d$ (S=1,I=0).  $qq\bar{b}\bar{b}$ states with  $J^P=0^+$ with (S=0,I=1) or with  $J^P=1^+$ with (S=1,I=1) was found to be unstable  against strong decay by BS~\cite{Brink:1998as}.  In the work by BS, the stability of the $T^1_{QQ}$ was obtained  from considering only the  $(q_1q_2)^{\bar{3}}\otimes(\bar{q}_3\bar{q}_4)^3$ component in the color wave function, without the color   $(q_1q_2)^{6}\otimes(\bar{q}_3\bar{q}_4)^{\bar{6}}$ component.
Thus, we are  committed to examining  the effect of the color $(q_1q_2)^{6}\otimes(\bar{q}_3\bar{q}_4)^{\bar{6}}$ component in Eq.~(\ref{spin1}) to the mass of the $T^1_{QQ}$.

Also, our work will  allow for possible couplings between the coordinates  $\pmb{\sigma}$, $\pmb{\lambda}$ and  $\pmb{\sigma^{\prime}}$ through the nonvanishing variational
parameters $C^s_{12}$, $C^s_{13}$, and $C^s_{23}$ appearing respectively in Eq.~(\ref{scheme1}) to Eq.~(\ref{scheme3}).  We will then compare our result to that of BS~\cite{Brink:1998as} using the Gaussian function in the absence of $C^s_{12}$, $C^s_{13}$ and $C^s_{23}$.

\section{Calculational schemes}

The total wave function must be antisymmetric under the transposition of (12) and (34) for $T^1_{QQ}$  because of the Pauli principle. Since we are interested in the lowest
orbital states with all quarks in the $l=0$ states, the spatial wave function should be symmetric.
Hence, the permutation property which should be satisfied by the color and spin part of the wave
function is symmetric under the transposition of (12) and antisymmetric under the transposition of
(34) because the flavor part of the wave functions is antisymmetric and symmetric for the light and
heavy quarks respectively. The above permutation properties only allow two states, $\psi_3$ and $\psi_4$ in Eq.~(\ref{spin1}).
\begin{align}
&(12)(q_1q_2)^{\bar{3}}_{0}\otimes(\bar{q}_3\bar{q}_4)^{3}_{1}=
 (q_1q_2)^{\bar{3}}_{0}\otimes(\bar{q}_3\bar{q}_4)^{3}_{1},  \nonumber \\
&(34)(q_1q_2)^{\bar{3}}_{0}\otimes(\bar{q}_3\bar{q}_4)^{3}_{1}=
 -(q_1q_2)^{\bar{3}}_{0}\otimes(\bar{q}_3\bar{q}_4)^{3}_{1}, \nonumber \\
&(12)(q_1q_2)^{6}_{1}\otimes(\bar{q}_3\bar{q}_4)^{\bar{6}}_{0}=
 (q_1q_2)^{6}_{1}\otimes(\bar{q}_3\bar{q}_4)^{\bar{6}}_{0},  \nonumber \\
&(34)(q_1q_2)^{6}_{1}\otimes(\bar{q}_3\bar{q}_4)^{\bar{6}}_{0}=
 -(q_1q_2)^{6}_{1}\otimes(\bar{q}_3\bar{q}_4)^{\bar{6}}_{0}.
\end{align}
The spatial function should therefore be symmetric under the transpositions.  We introduce different schemes depending on how this property is implemented. 

\subsection{Scheme 0}
The simplest way to implement the symmetry in the spatial wave function is to take  $C^s_{12}=C^s_{13}=C^s_{23}=0$ in the exponent of the Gaussian.
Considering only the variational parameters  $C^s_{11}$, $C^s_{22}$, and $C^s_{33}$ to be non-zero, the basis wave functions for $T_{QQ}^1$ can be written as the following:
\begin{equation}
\Psi^s_1=R^s(q_1q_2)^{\bar{3}}_{0}\otimes(\bar{q}_3\bar{q}_4)^{3}_{1},\quad
\Psi^s_2=R^s(q_1q_2)^{6}_{1}\otimes(\bar{q}_3\bar{q}_4)^{\bar{6}}_{0}. \label{base0}
\end{equation}
The spatial part of the basis wave functions in Eq.~(\ref{base0}) is given by excluding
$C^s_{12}$, $C^s_{13}$, and $C^s_{23}$ from Eq.~(\ref{space-wave-function}).

Scheme 0 :
\begin{equation}
 R^s=\exp[-(C^s_{11}\sigma^2+C^s_{22}\sigma^{\prime2}+C^s_{33}\lambda^2)].
\end{equation}
With this basis function, the Hamiltonian matrix  has the following form:
\begin{align}
\langle H \rangle=&\left(\begin{array}{cc}\sum m_i+\langle T_{cm} \rangle-\frac{3}{4}\langle V^C_1\rangle& 0 \\ 0 & \sum m_i+\langle T_{cm} \rangle-\frac{3}{4}\langle V^C_2 \rangle \end{array}\right)  \nonumber \\
&-\frac{3}{4}V^{SS}_{2\times2}, \label{Ham0}
\end{align}
where $\langle V^C_1 \rangle=-\frac{2}{3}(\langle V^C_{12} \rangle+\langle V^C_{34} \rangle)-\frac{1}{3}(\langle V^C_{13} \rangle+
\langle V^C_{14} \rangle+\langle V^C_{23} \rangle+\langle V^C_{24} \rangle)$ and $\langle V^C_2 \rangle=\frac{1}{3}(\langle V^C_{12} \rangle+\langle V^C_{34} \rangle)-\frac{5}{6}(\langle V^C_{13} \rangle+\langle V^C_{14} \rangle+\langle V^C_{23} \rangle+\langle V^C_{24} \rangle)$: The different sum in $\langle V^C_1 \rangle$ and $\langle V^C_2 \rangle$ come from different color wave function $\Psi^s_1$ and $\Psi^s_2$ respectively. The matrix element of the hyperfine potential $V^{SS}$ is given by
\begin{align}
&V^{SS}_{2\times2}|_{11}=2\langle V^{SS}_{12} \rangle-\frac{2}{3}\langle V^{SS}_{34} \rangle, \nonumber \\
&V^{SS}_{2\times2}|_{12}=-\frac{\sqrt{2}}{2}(\langle V^{SS}_{13} \rangle+\langle V^{SS}_{14} \rangle+\langle V^{SS}_{23} \rangle+\langle V^{SS}_{24} \rangle) \nonumber \\
&=V^{SS}_{2\times2}|_{21},\nonumber \\
&V^{SS}_{2\times2}|_{22}=\frac{2}{3}\langle V^{SS}_{12} \rangle-\langle V^{SS}_{34} \rangle.
\end{align}
The spatial part of the matrix element which was explained in detail by BS~\cite{Brink:1998as} can be obtained from the integration with respect to the three dimensional vector space which is illustrated by the three independent coordinate systems. The explicit forms are given in Eq.~(\ref{vcij}).

For the kinetic energy part, the kinetic operators in Eq.~(\ref{KE}) is given by
\begin{equation}
T=-\frac{\hbar^2}{2m_1}{\pmb{\nabla}}^2_{\sigma}
-\frac{\hbar^2}{2m_2}{\pmb{\nabla}}^2_{{\sigma}^{\prime}}
-\frac{\hbar^2}{2m^{\prime}}{\pmb{\nabla}}^2_{\lambda}.
\end{equation}
With this, the kinetic energy is given by
\begin{align}
\langle T_{cm} \rangle&=\langle R^s \vert T \vert R^s \rangle  \nonumber \\
&=\int d^3 \sigma d^3  {\sigma}^{\prime}
d^3 \lambda \exp(-Z^TC^sZ)T\exp(-Z^TC^sZ).
\end{align}
Similarly, the potential energy terms is
\begin{align}
\langle V^C_{ij} \rangle&=\langle R^s \vert (-\frac{\kappa}{r_{ij}}+\frac{r_{ij}}{a_0^2}-D) \vert R^s \rangle  \nonumber \\
&=\int d^3 \sigma d^3  {\sigma}^{\prime}
d^3 \lambda (-\frac{\kappa}{r_{ij}}+\frac{r_{ij}}{a_0^2}-D) \nonumber \\
& ~~~~~ \times \exp[-Z^T(C^s+C^s)Z] \label{vcij}
\end{align}
and
\begin{align}
\langle V^{SS}_{ij} \rangle=&\frac{\hbar^2c^2\kappa}{m_im_jc^4r_0^2}
\langle R^s \vert \frac{1}{r_{ij}}e^{-r_{ij}/r_0} \vert R^s \rangle  \nonumber \\
=&\frac{\hbar^2c^2\kappa}{m_im_jc^4r_0^2}
 \int d^3 \sigma d^3  {\sigma}^{\prime}d^3 \lambda \frac{1}{r_{ij}}e^{-r_{ij}/r_0}
 \nonumber \\
& ~~~~~ \times \exp[-Z^T(C^s+C^s)Z]. \label{vssij}
\end{align}
Depending on $r_{ij}$ appearing on the potential part, one needs to choose a convenient coordinate system among the three independent coordinate systems. This is easily done as the Jacobian related to coordinate transformation are all equal to one.   The calculation of these matrix element in terms of the color-spin states will be discussed in detail in the Appendix.

We have applied the variational method to the ground state using the basis set which is expressed
in scheme 0. In order to obtain the variational energy, we must minimize the lowest eigenvalue with
respect to the variational parameters after diagonalizing the matrix in Eq.~(\ref{Ham0}). Here, the variational
energy is obtained by differentiating the lowest eigenvalue with respect to the variational parameters. By analyzing the result in scheme 0, we first investigate the importance of the $R^s(q_1q_2)^{6}_{1}\otimes(\bar{q}_3\bar{q}_4)^{\bar{6}}_{0}$ component compared to $=R^s(q_1q_2)^{\bar{3}}_{0}\otimes(\bar{q}_3\bar{q}_4)^{3}_{1}$ component in the total wave function of $T^1_{bb}$.

\subsection{Other Schemes}

In these schemes, we hope to investigate the importance of introducing general  Gaussian wave functions to accommodate further correlations between quarks.  For that purpose, we introduce schemes I-V as below.  However, we only consider $(q_1q_2)^{\bar{3}}_{0}\otimes(\bar{q}_3\bar{q}_4)^{3}_{1}$ part as the color-spin basis function; because the contribution from the $(q_1q_2)^{6}_{1}\otimes(\bar{q}_3\bar{q}_4)^{\bar{6}}_{0}$ component is negligible as will be shown later through the analysis in scheme 0.
The schemes are introduced by adding the variational parameters $C^s_{ij}$ with $i \neq j$.

Scheme I :
\begin{align}
R^{s}_1=&\exp[-(C^s_{11}\sigma^2+C^s_{22}\sigma^{\prime2}+C^s_{33}\lambda^2+
2C^s_{12}\pmb{\sigma}\cdot\pmb{\sigma^{\prime}})]+ \nonumber \\
&\exp[-(C^s_{11}\sigma^2+C^s_{22}\sigma^{\prime2}+C^s_{33}\lambda^2-
2C^s_{12}\pmb{\sigma}\cdot\pmb{\sigma^{\prime}})]. \label{scheme1}
\end{align}

Scheme II :
\begin{align}
R^{s}_2=&\exp[-(C^s_{11}\sigma^2+C^s_{22}\sigma^{\prime2}+C^s_{33}\lambda^2+
2C^s_{13}\pmb{\sigma}\cdot\pmb{\lambda})]+ \nonumber \\
&\exp[-(C^s_{11}\sigma^2+C^s_{22}\sigma^{\prime2}+C^s_{33}\lambda^2-
2C^s_{13}\pmb{\sigma}\cdot\pmb{\lambda})].\label{scheme2}
\end{align}

Scheme III :
\begin{align}
R^{s}_3=&\exp[-(C^s_{11}\sigma^2+C^s_{22}\sigma^{\prime2}+C^s_{33}\lambda^2+
2C^s_{23}\pmb{\sigma^{\prime}}\cdot\pmb{\lambda})]+ \nonumber \\
&\exp[-(C^s_{11}\sigma^2+C^s_{22}\sigma^{\prime2}+C^s_{33}\lambda^2-
2C^s_{23}\pmb{\sigma^{\prime}}\cdot\pmb{\lambda})].\label{scheme3}
\end{align}
In the next scheme, we chose the parameters $C^s_{12}$, $C^s_{13}$ and $C^s_{23}$ determined from the previous scheme $i$ and then introduce a  linear combination of $R^s_1$, $R^s_2$, and $R^s_3$ with variational parameters $C_{i}$ for the wave function.

Scheme IV :
\begin{equation}
R^s=\sum_{i}C_{i}R^s_i
\end{equation}

Finally, we consider the more generalized spatial function with variational parameters $C^s_{13}\neq C^s_{23}\neq0$, and $C^s_{12}=0$.

Scheme V :
\begin{equation}
R^s_{(13)+(23)}=R^s_2+R^s_3
\end{equation}
We note that  all five spatial function in schemes I-V satisfies the symmetry  requirement under the transposition of (12) and (34).
\begin{table}[htdp]
\caption{The mass  and binding energy $B_T$ of $T^1_{bb}$ and $T^1_{bb}$ in scheme 0.    The units of mass and variational parameters are $~{\rm MeV}$, and $~{\rm fm}^{-2}$, respectively. The numbers in brackets are results when the color $(q_1q_2)^{6}_{1}\otimes(\bar{q}_3\bar{q}_4)^{\bar{6}}_{0}$ component is neglected. }
\begin{center}
\begin{tabular}{c|c|c|c}
\hline \hline
Type                  &    1Gaussian         & variational parameters                 &  $B_T$                    \\
\hline
$qq\bar{b}\bar{b}$ & 10576.6  & $C^s_{11}=2.8$, $C^s_{22}=18.4$,$C^s_{33}=2.7$ & -101.6                   \\
 &(10577.7) & ($C^s_{11}=2.9$, $C^s_{22}=18.5$,$C^s_{33}=2.9)$ &(-100.5) \\ \hline
$qq\bar{c}\bar{c}$ & 4036.4   & $C^s_{11}=2.7$, $C^s_{22}=6.9$,$C^s_{33}=2.5$   &  +97.9                   \\
 & (4043.9)& $(C^s_{11}=2.8$, $C^s_{22}=6.9$,$C^s_{33}=2.5)$ &(105.4) \\
\hline \hline
\end{tabular}
\end{center}
\label{tqq1-0-mass}
\end{table}

\subsection{Schemes with more Gaussian}

Finally, we investigate the importance of introducing correction to a simple Gaussian form.  This is simply accomplished by adding Gaussian with different overall coefficients.  To be specific, we first  introduce more Gaussian in the trial wave function
\begin{eqnarray}
\sum_jb_j\Psi^s_j \label{wave-more}
\end{eqnarray}
with variational parameters $b_j$ ($j=1 \sim 5$) and $\Psi^s_i=R^s_i(q_1q_2)^{\bar{3}}_{0}\otimes(\bar{q}_3\bar{q}_4)^{3}_{1}$.  With this, we will introduce the following additional schemes depending on how $R^{s}_i$ is defined.

Scheme VI :
\begin{align}
R^{s}_i=&\exp[-(C^{n_i}_{11}\sigma^2+C^{n_i}_{22}\sigma^{\prime2}+C^{n_i}_{33}
\lambda^2+2C^{n_i}_{13}\pmb{\sigma^{\prime}}\cdot\pmb{\lambda})]+ \nonumber \\
&\exp[-(C^{n_i}_{11}\sigma^2+C^{n_i}_{22}\sigma^{\prime2}+C^{n_i}_{33}\lambda^2-
2C^{n_i}_{13}\pmb{\sigma^{\prime}}\cdot\pmb{\lambda})].\label{scheme6}
\end{align}
The parameters of the Gaussian function, $C^{n_i}_{11}$, $C^{n_i}_{22}$, $C^{n_i}_{33}$, and $C^{n_i}_{13}$ are given by $C^{n_i}=\alpha^{n_i} C^0_{ij}$ where we chose $\alpha=$1.5, 2, 2.5\cite{Brink:1998as} and take $C^0_{ij}$ to be the variational parameters determined from the analysis in scheme II: $C^0_{11}=2.9~{\rm fm}^{-2}$, $C^0_{22}=18.5~{\rm fm}^{-2}$,
$C^0_{33}=2.9~{\rm fm}^{-2}$, and $C^0_{13}=0.6~{\rm fm}^{-2}$.
Then, for five Gaussian function, we take $n_i$ as the following:
\begin{equation}
n_1=-2, n_2=-1,n_3=0, n_4=1,n_5=2. \nonumber
\end{equation}

Scheme VII :
\begin{align}
R^{s}_i=&\exp[-(C^{n_i}_{11}\sigma^2+C^{n_i}_{22}\sigma^{\prime2}+C^{n_i}_{33}
\lambda^2+2C^{n_i}_{13}\pmb{\sigma}\cdot\pmb{\lambda})]+ \nonumber \\
&\exp[-(C^{n_i}_{11}\sigma^2+C^{n_i}_{22}\sigma^{\prime2}+C^{n_i}_{33}\lambda^2-
2C^{n_i}_{13}\pmb{\sigma}\cdot\pmb{\lambda})]+\nonumber \\
&\exp[-(C^{n_i}_{11}\sigma^2+C^{n_i}_{22}\sigma^{\prime2}+C^{n_i}_{33}
\lambda^2+2C^{n_i}_{23}\pmb{\sigma^{\prime}}\cdot\pmb{\lambda})]+ \nonumber \\
&\exp[-(C^{n_i}_{11}\sigma^2+C^{n_i}_{22}\sigma^{\prime2}+C^{n_i}_{33}\lambda^2-
2C^{n_i}_{23}\pmb{\sigma^{\prime}}\cdot\pmb{\lambda})].\label{scheme6}
\end{align}
Here, the parameters are defined in the same way as in scheme VI with $C^0_{ij}$ now taken from the analysis of scheme V.

The variational equations obtained by using the trial wave function in Eq.~(\ref{wave-more}) reduces to the following  eigenvalue problem with respect to $b_j$:
\begin{equation}
\sum_j\langle\Psi^s_i|H|\Psi^s_j\rangle b_j= \sum_jE\langle\Psi^s_i|\Psi^s_j
\rangle b_j.
\end{equation}

It should be noted that the trial wave function taken by BS with either single or five Gaussian did not take into account the correlations between quarks: $C^s_{ij}=C^{n_k}_{ij}=0$ for $i \neq j$.

\subsection{Normal meson}

In order to investigate the stability of $T^1_{QQ}$ against the decay into  a scalar and a vector meson, we calculated the mass of normal mesons using the Hamiltonian in Eq. (\ref{Hamiltonian}) with a two-body spatial function which was suggested by BS~\cite{Brink:1998as}. The spatial function has a form of Gaussian, given by,
\begin{equation}
\phi(r)=e^{-\frac{1}{2}a^2r^2}.\label{meson wave}
\end{equation}
where $r=(r_q-r_{\bar q})$ is the relative distance between quark and antiquark, and $a$ is a variational parameter. The list of the mass calculated by one Gaussian function is shown in Table \ref{normalmeson_mass}.
\begin{table}[htdp]
\caption{The masses of the pseudo-scalar and vector mesons containing a heavy antiquark  obtained using the variational method with one Gaussian function in Eq.~(\ref{meson wave}). The units of $a$ and masses are fm and MeV respectively. The experimental values are shown in the third line.
}
\begin{center}
\begin{tabular}{c|c|c|c|c}
\hline \hline
                                 &       $m_B$          &  $m_{B^*}$       &  $m_D$      &   $m_{D^*}$          \\
\hline
1Gaussian                & 5317              & 5360                & 1910        &  2028                \\
  $a^2$  &  8.81  & 8.07  &  8.96  & 7.30  \\
 experimental value    &   5279           &     5325                     & 1869          &     2006                         \\
\hline \hline
\end{tabular}
\end{center}
\label{normalmeson_mass}
\end{table}

\section{Numerical Results}

\subsection{Scheme 0}

We first analyze results in scheme 0 to investigate the importance of the color $6\bar{6}$ component in the total wave function of $T^1_{QQ}$.

In Table \ref{tqq1-0-mass}, we show the mass and variational parameters of $T^1_{bb}$ and $T^1_{bb}$ obtained from the coupled basis of $(q_1q_2)^{\bar{3}}_{0}\otimes(\bar{q}_3\bar{q}_4)^{3}_{1}$ and
$(q_1q_2)^{6}_{1}\otimes(\bar{q}_3\bar{q}_4)^{\bar{6}}_{0}$ in Scheme 0 with one Gaussian spatial function. Table \ref{tqq1-0-mass} also shows the binding energy $B_T=m_T-(M+M^{\prime})$ of the $T^1_{QQ}$ against the decay into a pseudo-scalar and a vector meson with mass  M and $M^{\prime}$, respectively. The masses of the mesons are calculated with the Hamiltonian in Eq.~(\ref{Hamiltonian}) using one variational Gaussian function.

As can be seen from table \ref{tqq1-0-mass}, the variational parameters changes little from those obtained by taking into account only the $(q_1q_2)^{\bar{3}}_{0}\otimes(\bar{q}_3\bar{q}_4)^{3}_{1}$ color component in the wave function.  This also  leads to only a small  change in the mass of 1$~{\rm MeV}$.
Hence, we can confirm that  the effect of the $(q_1q_2)^{6}_{1}\otimes(\bar{q}_3\bar{q}_4)^{\bar{6}}_{0}$ component on the binding of the heavy tetraquark system can be negligible.
As for the $T^1_{cc}$, we find that the mass of $T^1_{cc}$ is unbounded against strong decay.  In contrast to the case of $T^1_{bb}$, there is about 7 $~{\rm MeV}$ change in variational energy for the case of $T^1_{cc}$.

In scheme 0, the ground state which gives the lowest eigenvalue in Eq.(30) can be expressed as a linear combination of $\Psi_1$ and $\Psi_2$ in Eq.(28).  The mixing angle corresponding to the coefficients of $\Psi_1$ and $\Psi_2$ is shown in Table \ref{tqq1-0-wave}.  One can see that the mixing of the  $6\bar{6}$ component is again negligible.


\begin{table}[htdp]
\caption{The ground state wave function for $T^1_{QQ}$ in scheme 0.}
\begin{center}
\begin{tabular}{c|c}
\hline \hline
Type                  &    ground state                                   \\
\hline
$qq\bar{b}\bar{b}$   &  0.99671$\Psi_1$+0.080951$\Psi_2$             \\
$qq\bar{c}\bar{c}$ &    0.99558$\Psi_1$+0.093919$\Psi_2$              \\
\hline \hline
\end{tabular}
\end{center}
\label{tqq1-0-wave}
\end{table}

\subsection{Other Schemes}

Here, we present the results in the other schemes.
\begin{enumerate}

\item In scheme I, we find $C^s_{11}=2.9~{\rm fm}^{-2}$, $C^s_{22}=18.5~{\rm fm}^{-2}$,
$C^s_{33}=2.9~{\rm fm}^{-2}$, and $C^s_{12}=0.4~{\rm fm}^{-2}$. The values are the the same as in the case of $C^s_{12}=0$. The corresponding lowest energy is 10577.3$~{\rm MeV}$ which is also nearly equal to the value obtained with $C^s_{12}=0$. The correlation between $\sigma$ and ${\sigma}^{\prime}$ makes little difference on the structure of the spatial coordinate configuration.

\item Again, little change occurs for the case of scheme II describing the correlation between $\sigma$ and $\lambda$. The presence of $C^s_{13}$ lowers the ground state energy by 2 MeV with the values  $C^s_{11}=2.9~{\rm fm}^{-2}$, $C^s_{22}=18.5~{\rm fm}^{-2}$ , $C^s_{33}=2.9~{\rm fm}^{-2}$, and $C^s_{13}=0.6~{\rm fm}^{-2}$.

\item In scheme III, we find $C^s_{11}=2.9~{\rm fm}^{-2}$, $C^s_{22}=18.5~{\rm fm}^{-2}$,
$C^s_{33}=2.9~{\rm fm}^{-2}$, and $C^s_{23}=0.4~{\rm fm}^{-2}$ with the corresponding lowest energy 10577 MeV.

\item In scheme IV, we find the mass to be 10574.1$~{\rm MeV}$.

\item In scheme V, the variational parameters are given as $C^s_{11}=2.9~{\rm fm}^{-2}$, $C^s_{22}=18.5~{\rm fm}^{-2}$,
$C^s_{33}=2.9~{\rm fm}^{-2}$, $C^s_{13}=0.9~{\rm fm}^{-2}$, and
$C^s_{23}=0.6~{\rm fm}^{-2}$ with the lowest energy 10575.5 MeV.

\item In scheme VI,
we find that the lowest energy with five Gaussian functions with $\alpha=2$ to be 10558 MeV.

\item In scheme VII, we  find that the lowest energy with five Gaussian functions with $\alpha=2$ to be the same as that obtained from scheme VI.
\end{enumerate}

As can be seen from Table \ref{tetraquark1_mass}, we find from the analysis of  scheme I-VII, that our extended versions, taking into account correlations between quarks, did not give meaningful changes from the values obtained by BS~\cite{Brink:1998as} with either one Gaussian function or five Gaussian function without the correlations.   We also find that changing $\alpha$=1.5 and $\alpha$=2.5 do not introduce any additional changes, as was also noted  by BS~\cite{Brink:1998as}. Comparing the results from scheme I-V to those from scheme VI-VII, one finds that there is only a small change in the mass suggesting that   single Gaussian already encodes the dominant part of the total wave function. Moreover, the effect of including minimal correlation through scheme I to V induces even smaller mass change.  Hence, we omitted the variational calculation where more complicated correlation are present through
$C^s_{12}\neq C^s_{13}\neq C^s_{23}\neq 0$. Preliminary investigations suggests that this independence only persists when the antidiquarks are composed of heavy anti-quarks so that the system is intrinsically small.
We anticipate  that the dependence of $C^s_{12}$, $C^s_{13}$ and $C^s_{23}$ in lowering the variational energy is related to the size of the tetraquark to be considered.

\begin{table}[htdp]
\caption{ The mass of $T^1_{bb}$ in schemes I-VII.
}
\begin{center}
\begin{tabular}{c|c|c|c|c|c}
\hline \hline
$qq\bar{b}\bar{b}$  &  Sche-I  & Sche-II  &  Sche-III &  Sche-IV  & Sche-V \\
\hline
1Gaussian                  &   10577.3   &  10575.5         & 10577             & 10574.1         & 10575.5                          \\
\cline{2-6}
                     & \multicolumn{3}{|c}{Sche-VI} &  \multicolumn{2}{|c}{Sche-VII}
\\
5Gaussian                  &  \multicolumn{3}{|c}{10558}    &   \multicolumn{2}{|c}{10558}                          \\
\hline \hline
$qq\bar{b}\bar{b}$ &  \multicolumn{5}{|c}{Brink-Stancu }      \\
\hline
1Gaussian                &  \multicolumn{5}{|c}{10577.7}        \\
5Gaussian                &  \multicolumn{5}{|c}{10558.1}          \\
\hline \hline
\end{tabular}
\end{center}
\label{tetraquark1_mass}
\end{table}

\begin{table}[htdp]
\caption{ The mass of $T^1_{cc}$ without the component $(q_1q_2)^{6}_{1}\otimes(\bar{q}_3\bar{q}_4)^{\bar{6}}_{0}$ in the color-spin space.
}
\begin{tabular}{c|c|c}
\hline\hline
  $qq\bar{c}\bar{c}$   & Sche-0   & Sche-II \\
\hline
1Gaussian                  & 4043.9         &  4042.7     \\
\hline\hline
\end{tabular}
\label{tetraquark1cc_mass}
\end{table}

This effect is also true for $T^1_{cc}$ as the mass change only by 1 MeV as can be seem from Table \ref{tetraquark1cc_mass}.
In obtaining the values for Table \ref{tetraquark1cc_mass}, we only took into account the $(q_1q_2)^{\bar{3}}_{0}\otimes(\bar{q}_3\bar{q}_4)^{3}_{1}$ color component for the trial wave function without correlation and with a minimal correlation as given in scheme II.

\subsection{Sizes of hadrons}

It is useful to look at the relative distances between quarks in each hadron.  From table \ref{normalmeson_mass}, we note that the distance between the quark and antiquark in the $B$ meson is $r_B \sim \sqrt{2}a^{-1} =0.476 $ fm; similar values are obtained for $B^*,D$ and $D^*$ mesons.  For $T^1_{bb}$ meson, the distance between the diquark and antidiquark is $r_\lambda \sim 1/\sqrt{C_{33}} =0.608 $ fm and that between the $b$ quarks is $r_{\sigma'} \sim \sqrt{2/C_{22}} =0.329$ fm.  For  $T^1_{cc}$, while $r_\lambda \sim 1/\sqrt{C_{33}} =0.632 $ fm is similar to that of $T^1_{bb}$, $r_{\sigma'} \sim \sqrt{2/C_{22}} =0.530$ fm is much larger.

\section{The mass splitting in Hyperfine potential}

In this section, we investigate the contribution of the hyperfine potential term which is crucial for deciding the stability against strong decay. In particular, we perform two calculations.  In the first part, we calculate the contribution of the hyperfine potential within Scheme 0 of our variational method.  In a second approach, we estimate these from  fitting it to the mass differences between the mesons and baryons with constant factors.   Let us elaborate on the second approach.  We  introduce $C_{ij}$, which should be not confused with the variational parameters $C^s_{ij}$, for the following parametrization to the mass coming from the hyperfine potential:
\begin{eqnarray}
V^{SS}=-\sum_{i<j}C_{ij}\lambda^c_i\lambda^c_j{\sigma}_i\cdot{\sigma}_j.
\label{interaction}
\end{eqnarray}
In the first estimate, $C_{ij}= \langle V^{SS}_{ij} \rangle $ as given in Eq.~(\ref{vssij}), and can be calculated within variational approach. In the second approach, we assume that $C_{ij}$ depends only on the flavor and whether the pair is a quark-quark or quark-antiquark type. 
Then, $C_{ij}$ 
can be extracted from the observed mass differences between the baryons or mesons, within the constituent quark model. Our purpose is to assess whether one can determine the stability of tetraquark states by looking at only the  hyperfine potential term given in Eq.~(\ref{interaction}) and assumptions within our second approach. For a meson consisting of a quark and antiquark, the contribution of the color part to the interaction Hamiltonian in Eq.~(\ref{interaction}) is -16/3, and the spin part is either -3 or 1 for S=0 and S=1, respectively.
From the mass differences $J/\psi - \eta_c, D^*-D, \rho-\pi, B^*-B,\Upsilon-\eta_b$~\cite{Beringer:1900zz} we find
\begin{center}
\begin{tabular}{c}
      $C_{u\bar{c}}=C_{d\bar{c}}=6.7~{\rm MeV}$,\qquad$C_{u\bar{b}}=C_{d\bar{b}}=2.2~{\rm MeV}$\\
 $C_{u\bar{u}}=29.5~{\rm MeV}$,\quad  $C_{c\bar{c}}=5.48~{\rm MeV}$,
 \quad  $C_{b\bar{b}}=3.25~{\rm MeV}$.
\end{tabular}
\end{center}
\begin{table}
\caption{The expectation value of interaction Hamiltonian in Eq.~(\ref{interaction}) for mesons and baryons is shown below. For baryons with S=1/2, the configuration of flavors is $qqq^{\prime}$ including two identical quarks.}
\begin{tabular}{c|c|c}
\hline\hline
  & S=0& S=1\\
\cline{2-3}
Meson &-$16C_{12}$ & $\frac{16}{3}C_{12}$\\
\hline
 & S=1/2 & S=3/2 \\
\cline{2-3}
 Baryon&$\frac{8}{3}(C_{12}-4C_{13})$&$\frac{8}{3}(C_{12}+C_{13}+C_{23})$\\
 \hline\hline
\end{tabular}
\label{meson-vss}
\end{table}

For baryons, the expectation value of the color operator $\lambda^c_i\lambda^c_j$ with respect to a color singlet wave function $\epsilon^{ijk}q_i(1)q_j(2)q_k(3)$ is the same for all the pairs and equal to -8/3.  Specifically, $\langle V^{SS} \rangle= 8/3(C_{12}+C_{23}+C_{13}$) for S=3/2 baryons, and $\langle V^{SS} \rangle = 8/3(C_{12}-4C_{13}$) for S=1/2 baryons when two quarks are identical $q$$q$$q^\prime$~\cite{Hogaasen:2007nw}. From the nucleon $\Delta$ and $N$ mass difference, we have
\begin{center}
\begin{tabular}{c}
$C_{uu}=C_{ud}=18~{\rm MeV}$\\
\end{tabular}
\end{center}

On the other hand, the strength factor involving  two heavy quarks such as $C_{cc}$ can be inferred from the value of $C_{c\bar{c}}$ with the same ratio as in the case of light quarks  $C_{u\bar{u}}=1.63 C_{uu}$ as can be seen in our estimation:  we will assume  $C_{cc}=1/1.63C_{c\bar{c}}$ and $C_{bb}=1/1.63C_{b\bar{b}}$. Then we have :
\begin{center}
\begin{tabular}{c}
$C_{cc}=3.36~{\rm MeV}$\quad \quad  $C_{bb}=1.99~{\rm MeV}$.
\end{tabular}
\end{center}

Now, to calculate the hyperfine splitting within our second approach, we note that the matrix element of the hyperfine potential for $T^1_{bb}$ and $T^1_{cc}$ in terms of $(q_1q_2)^{\bar{3}}_{0}\otimes(\bar{q}_3\bar{q}_4)^{3}_{1}$ and $(q_1q_2)^{6}_{1}\otimes(\bar{q}_3\bar{q}_4)^{\bar{6}}_{0}$ is written as
\begin{equation}
\langle V^{SS} \rangle=-2/3\left(\begin{array}{cc}  V^{SS}_{11} &  V^{SS}_{12} \\
  V^{SS}_{21} &  V^{SS}_{22} \end{array} \right)\nonumber
\end{equation}
with
\begin{align}
 &V^{SS}_{11}=12C_{12}-4C_{34},\nonumber\\
&V^{SS}_{12}=V^{SS}_{21}=-3\sqrt{2}(C_{13}+C_{14}+C_{23}+C_{24}),\nonumber\\
&V^{SS}_{22}=2C_{12}-6C_{34}.\label{matrix element}
\end{align}
In the second approach, we use the phenomenological estimates in the right hand side of Eq.~(\ref{matrix element}).  The final values are given in the last (4'th) column of table \ref{tetraquark4_mass}.

\begin{table}[htdp]
\caption{ The list of the value of each term of the Hamiltonian in Eq.(1) for $T^1_{bb}$ and $T^1_{cc}$  : To compare the mass splitting in the hyperfine potential
in Eq.~(\ref{interaction}) with that of Eq.(31), the lowest eigenvalue in Eq.~(\ref{interaction}) is putted in column 4
}
\begin{center}
\begin{tabular}{c|c|c|c}
\hline \hline
   Type                      &           $H_0$            &  $V^{SS}$          &                       $V^{SS}$                   \\
\hline
 $qq\bar{b}\bar{b}$    &       10756              &      -181.4          &             -143.5                                   \\
$B$                                   &        5351.4            &        -34.0            &           -35.2                                     \\
$B^*$                                &        5350.4            &         10.5            &            11.7                                      \\
\cline{2-4}
             &  $H_0-H^{M+{M^{\prime}}}_0$ & $V^{SS}-V^{SS}_{M+{M^{\prime}}}$
             &   $V^{SS}-V^{SS}_{M+{M^{\prime}}}$ \\
                                          &  54.5                     &           -157.9                &      -120            \\
\hline
   Type             &           $H_0$            &  $V^{SS}$          &                       $V^{SS}$            \\
\hline
$qq\bar{c}\bar{c}$     &         4215.1            &        -186.9          &            -170.8                                 \\
$D$                                   &          2007.8           &         -97.2         &            -107.2                                 \\
$D^*$                                &          2000.9           &          27.1         &              35.5                                  \\
\cline{2-4}
             &  $H_0-H^{M+{M^{\prime}}}_0$ &  $V^{SS}-V^{SS}_{M+{M^{\prime}}}$
             &  $V^{SS}-V^{SS}_{M+{M^{\prime}}}$\\
                                        &  206.4                  &     -116.8                &        -99.1                                      \\
\hline \hline
\end{tabular}
\end{center}
\label{tetraquark4_mass}
\end{table}

In Table \ref{tetraquark4_mass}, we also show the value of each part of the Hamiltonian in Eq.~(\ref{Hamiltonian}) calculated within scheme 0.  $H_0$ corresponds to the kinetic and confinement potential terms calculated in scheme 0 for $T^1_{QQ}$. $H^{M+{M^{\prime}}}_0$ are the corresponding sum for the scalar and  vector meson using the Gaussian function in Eq.~(\ref{meson wave}). $V^{SS}$ in column 3 represents the eigenvalue of the matrix of hyperfine potential in terms of the basis set in Eq.(28) for the heavy tetraquark. The $V^{SS}_{M+{M^{\prime}}}$  are the values of a scalar and a vector meson for hyperfine potential with one Gaussian function in Eq.~(\ref{meson wave}).

As shown in the Table \ref{tetraquark4_mass}, the difference of $H_0$ between the heavy tetraquark and the sum of a scalar and a vector meson becomes considerably smaller in $T^1_{bb}$ than in $T^1_{cc}$.  As can be seen from table \ref{tqq1-0-mass}, the main difference between these two tetraquarks is in the average distance between the two heavy antiquarks.   When the heavy antiquark becomes large, we can estimate the binding energy simply by looking at the difference of hyperfine potential;  that is, $V^{SS}-V^{SS}_{M+{M^{\prime}}}$ in column 4, provided that $H_0-(H^M_0+H^{M^{\prime}}_0)=0$ for $T^1_{bb}$.

\section{Summary}

With a simple variational Gaussian function, which is convenient to analyze the states with L=0 configurations, we have calculated the ground state  energy of the $J^P=1^+$  $ud\bar{b}\bar{b}$ tetraquark containing two identical heavy antiquarks in a nonrelativistic potential model with color confinement and spin hyperfine interaction.   In particular, we extend the the work by BS to investigate the effect of including the color anti-sextet component of the  diquark configuration as well as using several more Gaussian parametrization for the spatial wave function.
From the analysis in Scheme 0, we find that taking into account the $R^s(q_1q_2)^{6}_{1}\otimes(\bar{q}_3\bar{q}_4)^{\bar{6}}_{0}$ has little effect on the binding as well as on the wave function of the tetraquark state, whose wave function is dominated by the $R^s(q_1q_2)^{\bar{3}}_{0}\otimes(\bar{q}_3\bar{q}_4)^{3}_{1}$ component, as was expected by BS~\cite{Brink:1998as}.

For  the heavy tetraquark, we also find that the variational energy does not depend very much on whether we allow for the nonvanishing parameters $C^s_{12}\neq C^s_{13}\neq C^s_{23}\neq 0$ or $C^s_{12}=C^s_{13}=C^s_{23}=0$ in the exponent in Eq.(9). Still, we find that the inclusion of variational parameter 
$C^s_{13}$ introduces the most important change in the mass.   This suggests that the orientation of the heavy antiquark $\sigma'$ is relatively less important compared to the other orientations involving light quarks.  Therefore, we expect that this nonvanishing variational parameters might play a more important role in the light tetraquark system.  Finally, in section VI, we have shown that the mass splitting of hyperfine potential can provide an intuitive picture for the binding energy of $T^1_{bb}$ against the $B,B^*$.

We still find that the $T^1_{bb}$ mass we obtain, which is consistent with that  by BS~\cite{Brink:1998as},  remains about 33 $~{\rm MeV}$ higher than that obtained   by Silvestre-Brac and Semay~\cite{SilvestreBrac:1993ry} using a harmonic oscillator basis with the same Hamiltonian. A possible further improvement in our calculation is to take into account the coupling to the asymptotic decay channels which is appropriate for describing the decay property as was argued  by BS~\cite{Brink:1998as}.  Moreover, although we have neglected the center of mass motion for all mesons, these might not cancel between the tetraquark and two meson sates.  Also, we have assumed that the constant $D$ in Eq.~(\ref{vc_ij}) is univeral for both the tetraquark and meson.  All these issues remains to be a caveat in our approach that should be address later.

\appendix

\section{Color-singlet states for tetraquark}
In this section, we will calculate the matrix of the interaction Hamiltinian in terms of the color-spin wave function which have been introduced in the previous section. It is essential to mention the Casimir operator of $SU(3)_c$ for the purpose of investigating the action of $\lambda^c_i\lambda^c_j$ on the color singlet. According to $Schur^,s$ lemmas, the Casimir operator, $\lambda^c\lambda^c$ can be expressed as a multiple of the unit matrix in any irreducible representation of $SU(3)_c$ because the Casimir operator commutes with all of the irreducible representation of $SU(3)_c$. Therefore, each basis vector belonging to a multiplet of any irreducible representation has a common eigenvalue to the Casimir operator. In addition, $SU(3)$ has a second invariant operator, whose the eigenvalues also characterize the multiplets of  $SU(3)$. Then, $Racah^,s$ theorem tells us that with the two invariant operator, the $SU(3)$ multiplets are completely classified. There are several kinds of irreducible representation related to $SU(3)$. :\\
\begin{tabular}{c}
[1]=D(0,0)
\end{tabular}
 \begin{tabular}{c}
[3]=D(1,0)=
\end{tabular}
\begin{tabular}{|c|}
 \hline
$\quad$ \\
\hline
\end{tabular}
\qquad
\begin{tabular}{c}
$[\bar{3}]$=D(0,1)=
\end{tabular}
\begin{tabular}{|c|}
\hline
  $\quad$  \\
\cline{1-1}  $\quad$   \\
\hline
\end{tabular}
\\
\begin{tabular}{c}
[6]=D(2,0)=
\end{tabular}
\begin{tabular}{|c|c|}
\hline
 $\quad$&$\quad$    \\
\hline
\end{tabular}
\quad
\begin{tabular}{c}
$[\bar{6}]$=D(0,2)=
\end{tabular}
\begin{tabular}{|c|c|}
\hline
$\quad$ & $\quad$   \\
\cline{1-2}$\quad$ &  $\quad$  \\
\hline
\end{tabular}
\\
In the irreducible representation D($p_1$,$p_2$), the number $p_k$ appearing in the $"k"$'th position denotes the number of columns with k boxes in a given Young diagram. We define the multiplets of $SU(3)$ as $\psi{(D(p_1,p_2))}$.  Then, the action of $\lambda^c\lambda^c$ on $\psi{(D(p_1,p_2))}$ gives the well-known formula
\begin{align}
\lambda^c\lambda^c&\psi{(D(p_1,p_2))}=\nonumber\\
&\frac{4}{3}(p_1^2+p_1p_2+p_2^2+3p_1+3p_2)\psi{(D(p_1,p_2))}.
\end{align}
For example, we have :
\begin{align}
&\lambda^c\lambda^c\psi{(D(0,0))}=0,\nonumber\\
&\lambda^c\lambda^c\psi{(D(1,0))}=16/3\psi{(D(1,0))},\nonumber\\
&\lambda^c\lambda^c\psi{(D(0,1))}=16/3\psi{(D(0,1))},\nonumber\\
&\lambda^c\lambda^c\psi{(D(2,0))}=40/3\psi{(D(2,0))},\nonumber\\
&\lambda^c\lambda^c\psi{(D(0,2))}=40/3\psi{(D(0,2))},\nonumber\\
&\lambda^c\lambda^c\psi{(D(1,1))}=12\psi{(D(1,1))}.\nonumber\\
\end{align}
In order to calculate the matrix element of $\lambda^c_i\lambda^c_j$ with respect to the muliplet of  $SU(3)_c$ in tetraquark, we need to descibe two color singlets coming from a singlet-singlet color and an octet-octet color state. We denote two color singlets by $(q_1\bar{q_3})^{c=1}\otimes(q_2\bar{q}_4)^{c=1},
(q_1\bar{q_3})^{c=8}\otimes(q_2\bar{q}_4)^{c=8}$ or $(q_1\bar{q_4})^{c=1}\otimes(q_2\bar{q}_3)^{c=1},
(q_1\bar{q_4})^{c=8}\otimes(q_2\bar{q}_3)^{c=8}$. We can find two color singlets with a irreducible tensor methods :
\begin{equation}
(q_1\bar{q_3})^{1}\otimes(q_2\bar{q}_4)^{1}=\frac{1}{3}q^i(1)\bar{q}_i(3)
q^j(2)\bar{q}_j(4),\nonumber\\
\end{equation}
\begin{align}
(q_1\bar{q_3})^{8}\otimes(q_2\bar{q}_4)^{8}=&\frac{1}{2\sqrt{2}}(q^i(1)\bar{q}_j(3)\nonumber\\
&-\frac{1}{3}\delta^i_jq^k(1)\bar{q}_k(3))(q^j(2)\bar{q}_i(4)\nonumber\\
&-\frac{1}{3}\delta^j_iq^k(2)\bar{q}_k(4)).
\end{align}
where $q^i(1)\bar{q}_j(3)
-\frac{1}{3}\delta^i_jq^k(1)\bar{q}_k(3)$ indicates the irreducible tensor of octet multiplet. It is easy to see that these color singlets are orthogonal to each other. Hence, a two dimensional vector space is spanned by $(q_1\bar{q_3})^{1}\otimes(q_2\bar{q}_4)^{1}$, and $(q_1\bar{q_3})^{8}\otimes(q_2\bar{q}_4)^{8}$. For the same reason, $ (q_1q_2)^{\bar{3}}\otimes(\bar{q}_3\bar{q}_4)^3$ and $(q_1q_2)^{6}\otimes(\bar{q}_3\bar{q}_4)^{\bar{6}}$ which are orthogonal constitute the identical two dimensional vector space. There exists uniquely an isomorphism which is called an one-to-one correspondence such that the transformation from one bases to the other is an orthogonal 2 by 2 matrix because of the conservation of inner product. The transformation is given by,
\begin{align}
  &(q_1\bar{q_3})^{1}\otimes(q_2\bar{q}_4)^{1}=\nonumber\\
  &\frac{1}{\sqrt{3}}(q_1q_2)^{\bar{3}}\otimes(\bar{q}_3\bar{q}_4)^3+
  \sqrt{\frac{2}{3}}(q_1q_2)^{6}\otimes(\bar{q}_3\bar{q}_4)^{\bar{6}},\nonumber\\
  &(q_1\bar{q_3})^{8}\otimes(q_2\bar{q}_4)^{8}=\nonumber\\
  &-\sqrt{\frac{2}{3}}(q_1q_2)^{\bar{3}}\otimes(\bar{q}_3\bar{q}_4)^3+
  \frac{1}{\sqrt{3}}(q_1q_2)^{6}\otimes(\bar{q}_3\bar{q}_4)^{\bar{6}}.
 \end{align}
We can also find the transformation from  the basis set of $(q_1\bar{q_4})^{1}\otimes(q_2\bar{q}_3)^{1}$ and
$(q_1\bar{q_4})^{8}\otimes(q_2\bar{q}_3)^{8}$ to the basis set of  $(q_1q_2)^{\bar{3}}\otimes(\bar{q}_3\bar{q}_4)^3$ and  $(q_1q_2)^{6}\otimes(\bar{q}_3\bar{q}_4)^{\bar{6}}$ :
\begin{align}
  &(q_1\bar{q_4})^{1}\otimes(q_2\bar{q}_3)^{1}=\nonumber\\
  &-\frac{1}{\sqrt{3}}(q_1q_2)^{\bar{3}}\otimes(\bar{q}_3\bar{q}_4)^3+
  \sqrt{\frac{2}{3}}(q_1q_2)^{6}\otimes(\bar{q}_3\bar{q}_4)^{\bar{6}},\nonumber\\
  &(q_1\bar{q_4})^{8}\otimes(q_2\bar{q}_3)^{8}=\nonumber\\
  &\sqrt{\frac{2}{3}}(q_1q_2)^{\bar{3}}\otimes(\bar{q}_3\bar{q}_4)^3+
  \frac{1}{\sqrt{3}}(q_1q_2)^{6}\otimes(\bar{q}_3\bar{q}_4)^{\bar{6}}.
 \end{align}
We are now in a position to apply $\lambda^c_i\lambda^c_j$ on the color singlets. For a system consisting of two quarks $i$ and $j$, the generators are $ \lambda^c_{ij}= \lambda^c_i+\lambda^c_j$, where c runs from 1 to 8. Then, by analogy with angular momentum, $ \lambda^c_i \lambda^c_j$ can be expressed as,
\begin{equation}
\lambda^c_i \lambda^c_j=\frac{1}{2}((\lambda^c_{ij})^2-(\lambda^c_i)^2-(\lambda^c_j)^2),
\end{equation}
where $(\lambda^c_{ij})^2$, $(\lambda^c_i)^2$ and $(\lambda^c_j)^2$ are Casimir operators associated with the two-body system, and the particles $i$ and $j$, respectively. Applying $\lambda^c_1\lambda^c_2$ to $(q_1q_2)^{\bar{3}}\otimes(\bar{q}_3\bar{q}_4)^3$ gives,
\begin{align}
\lambda^c_1\lambda^c_2(q_1q_2)^{\bar{3}}\otimes(\bar{q}_3\bar{q}_4)^3&=
\frac{1}{2}((\lambda^c_{12})^2(q_1q_2)^{\bar{3}})\otimes(\bar{q}_3\bar{q}_4)^3
\nonumber\\
&-(((\lambda^c_1)^2q_1)q_2)^{\bar{3}})\otimes(\bar{q}_3\bar{q}_4)^3\nonumber\\
&-(q_1((\lambda^c_{2})^2q_2))^{\bar{3}}\otimes(\bar{q}_3\bar{q}_4)^3 \nonumber\\
&=\frac{1}{2}(\frac{16}{3}-\frac{16}{3}-\frac{16}{3})(q_1q_2)^{\bar{3}}\otimes
(\bar{q}_3\bar{q}_4)^3 \nonumber\\
&=-\frac{8}{3}(q_1q_2)^{\bar{3}}\otimes(\bar{q}_3\bar{q}_4)^3.
\end{align}
Similarly, applying $\lambda^c_1\lambda^c_2$ on  $(q_1q_2)^{6}\otimes(\bar{q}_3\bar{q}_4)^{\bar{6}}$ yields,
\begin{align}
\lambda^c_1\lambda^c_2(q_1q_2)^{6}\otimes(\bar{q}_3\bar{q}_4)^{\bar{6}}&=
\frac{1}{2}((\lambda^c_{12})^2(q_1q_2)^{6})\otimes(\bar{q}_3\bar{q}_4)^{\bar{6}}
\nonumber\\
&-(((\lambda^c_1)^2q_1)q_2)^{6})\otimes(\bar{q}_3\bar{q}_4)^{\bar{6}}\nonumber\\
&-(q_1((\lambda^c_{2})^2q_2))^{6})\otimes(\bar{q}_3\bar{q}_4)^{\bar{6}}\nonumber\\
&=\frac{1}{2}(\frac{40}{3}-\frac{16}{3}-\frac{16}{3})(q_1q_2)^{6}\otimes
(\bar{q}_3\bar{q}_4)^{\bar{6}}\nonumber\\
&=+\frac{4}{3}(q_1q_2)^{6}\otimes(\bar{q}_3\bar{q}_4)^{\bar{6}}.
\end{align}
It follows immediately that the same result is obtained for the $\lambda^c_3\lambda^c_4$. For the operator, $\lambda^c_1\lambda^c_3$, the basis set of  $(q_1\bar{q_3})^{1}\otimes(q_2\bar{q}_4)^{1}$ and
$(q_1\bar{q_3})^{8}\otimes(q_2\bar{q}_4)^{8}$ instead of the basis set of $ (q_1q_2)^{\bar{3}}\otimes(\bar{q}_3\bar{q}_4)^3$ and $(q_1q_2)^{6}\otimes(\bar{q}_3\bar{q}_4)^{\bar{6}}$ is required to calculate the matrix element.  Then, the matrix element of $\lambda^c_1\lambda^c_3$, in terms of  $ (q_1q_2)^{\bar{3}}\otimes(\bar{q}_3\bar{q}_4)^3$ and $(q_1q_2)^{6}\otimes(\bar{q}_3\bar{q}_4)^{\bar{6}}$, is obtained from the similarity transformation which changes the matrix representation based on a basis set. In a similar way, we have :
\begin{align}
\lambda^c_1\lambda^c_3(q_1\bar{q_3})^{1}\otimes(q_2\bar{q}_4)^{1}&=
\frac{1}{2}(((\lambda^c_{13})^2(q_1\bar{q_3})^{1})\otimes(q_2\bar{q}_4)^{1}\nonumber\\
&-(((\lambda^c_1)^2q_1)\bar{q_3})^{1})\otimes(q_2\bar{q}_4)^{1}\nonumber\\
&-(q_1((\lambda^c_{3})^2\bar{q_3}))^{1}\otimes(q_2\bar{q}_4)^{1} \nonumber\\
&=\frac{1}{2}(0-\frac{16}{3}-\frac{16}{3})(q_1\bar{q_3})^{1}\otimes(q_2\bar{q}_4)^{1} \nonumber\\
&=-\frac{16}{3}(q_1\bar{q_3})^{1}\otimes(q_2\bar{q}_4)^{1}.
\end{align}
\begin{align}
\lambda^c_1\lambda^c_3(q_1\bar{q_3})^{8}\otimes(q_2\bar{q}_4)^{8}&=
\frac{1}{2}(((\lambda^c_{13})^2(q_1\bar{q_3})^{8})\otimes(q_2\bar{q}_4)^{8}\nonumber\\
&-(((\lambda^c_1)^2q_1)\bar{q_3})^{8})\otimes(q_2\bar{q}_4)^{8}\nonumber\\
&-(q_1((\lambda^c_{3})^2\bar{q_3}))^{8}\otimes(q_2\bar{q}_4)^{8} \nonumber\\
&=\frac{1}{2}(12-\frac{16}{3}-\frac{16}{3})(q_1\bar{q_3})^{8}\otimes(q_2\bar{q}_4)^{8} \nonumber\\
&=+\frac{2}{3}(q_1\bar{q_3})^{8}\otimes(q_2\bar{q}_4)^{8}.
\end{align}
To calculate  the matrix element of $\lambda^c_1\lambda^c_3$ in terms of  $(q_1q_2)^{6}\otimes(\bar{q}_3\bar{q}_4)^{\bar{6}}$ and $(q_1q_2)^{\bar{3}}\otimes(\bar{q}_3\bar{q}_4)^3$, we find the inverse of transformation,
which is equivalent to the orthogonal matrix U, and
$\langle\lambda^c_1\lambda^c_3\rangle_{(q_1\bar{q_3})^{8}\otimes
(q_2\bar{q}_4)^{8},(q_1\bar{q_3})^{1}
\otimes(q_2\bar{q}_4)^{1}}$,
\begin{equation}
U=\left(\begin{array}{cc}\frac{1}{\sqrt{3}} &-\sqrt{\frac{2}{3}}\\  \sqrt{\frac{2}{3}}& \frac{1}{\sqrt{3}} \end{array} \right),\nonumber\\
\end{equation}

\begin{align}
\langle \lambda^c_1\lambda^c_3\rangle_{(q_1\bar{q_3})^{8}\otimes(q_2\bar{q}_4)^{8}, (q_1\bar{q_3})^{1}\otimes(q_2\bar{q}_4)^{1}}=&\nonumber\\
&\left(\begin{array}{cc}\frac{2}{3} &0\\  0& -\frac{16}{3} \end{array} \right)\nonumber\\.
\end{align}

Finally, we reach the matrix representation based on $(q_1q_2)^{6}\otimes(\bar{q}_3\bar{q}_4)^{\bar{6}}$ and $ (q_1q_2)^{\bar{3}}\otimes(\bar{q}_3\bar{q}_4)^3$ as,
\begin{align}
&\langle\lambda^c_1\lambda^c_3\rangle_{(q_1q_2)^{6}\otimes
(\bar{q}_3\bar{q}_4)^{\bar{6}},
(q_1q_2)^{\bar{3}}\otimes(\bar{q}_3\bar{q}_4)^3}\nonumber\\
&=U^T\langle \lambda^c_1\lambda^c_3\rangle_{(q_1\bar{q_3})^{8}\otimes(q_2\bar{q}_4)^{8},
(q_1\bar{q_3})^{1}\otimes(q_2\bar{q}_4)^{1}}U\nonumber\\
&=\left(\begin{array}{cc}\frac{1}{\sqrt{3}} &\sqrt{\frac{2}{3}}\\  -\sqrt{\frac{2}{3}}& \frac{1}{\sqrt{3}} \end{array} \right)\left(\begin{array}{cc}\frac{2}{3} &0\\  0& -\frac{16}{3} \end{array} \right)\left(\begin{array}{cc}\frac{1}{\sqrt{3}} &-\sqrt{\frac{2}{3}}\\ \sqrt{\frac{2}{3}}& \frac{1}{\sqrt{3}} \end{array} \right)\nonumber\\
&=\left(\begin{array}{cc}-\frac{10}{3} &-2\sqrt{2}\\ -2\sqrt{2} &-\frac{4}{3}\end{array} \right).
\end{align}

The basis set of $(q_1\bar{q_3})^{8}\otimes(q_2\bar{q}_4)^{8}$ and
$(q_1\bar{q_3})^{1}\otimes(q_2\bar{q}_4)^{1}$ is necessary to obtain the matrix element of $\lambda^c_1\lambda^c_4$.
\begin{table}
\caption{The matrix of $\lambda^c_i\lambda^c_j$ is written in terms of two basis set, $\phi_1=
(q_1q_2)^{6}\otimes(\bar{q}_3\bar{q}_4)^{\bar{6}}$ and $\phi_2=(q_1q_2)^{\bar{3}}\otimes(\bar{q}_3\bar{q}_4)^3$, $\psi_1=(q_1\bar{q_3})^{8}\otimes(q_2\bar{q}_4)^{8}$ and $\psi_2=(q_1\bar{q_3})^{1}\otimes(q_2\bar{q}_4)^{1}$.}
\begin{tabular}{|c|c|c|}
\hline\hline
& $\scriptscriptstyle{(q_1q_2)^{6}\otimes(\bar{q}_3\bar{q}_4)^{\bar{6}},(q_1q_2)^{\bar{3}}
\otimes (\bar{q}_3\bar{q}_4)^3}$& $\scriptscriptstyle{(q_1\bar{q_3})^{8}\otimes(q_2\bar{q}_4)^{8},(q_1\bar{q_3})^{1}
\otimes(q_2\bar{q}_4)^{1}}$  \\
\cline{1-3}  $\begin{array}{c}\scriptstyle{\lambda^c_1\lambda^c_2} \\
\scriptstyle{=\lambda^c_3\lambda^c_4} \end{array} $&  $\Big{(}\begin{array}{cc}\scriptstyle{\frac{4}{3}} & \scriptstyle{0} \\ \scriptstyle{0} &\scriptstyle{-\frac{8}{3}}\end{array} \Big{)}$ &  $\Big{(}\begin{array}{cc}\scriptstyle{-\frac{4}{3}} & \scriptstyle{\frac{4\sqrt{2}}{3}} \\ \scriptstyle{\frac{4\sqrt{2}}{3}} &\scriptstyle{0}\end{array} \Big{)}$    \\
\cline{1-3} $\begin{array}{c}\scriptstyle{\lambda^c_1\lambda^c_3} \\ \scriptstyle{=\lambda^c_2\lambda^c_4} \end{array} $ &  $\Big{(}\begin{array}{cc}\scriptstyle{-\frac{10}{3}} & \scriptstyle{-2\sqrt{2}} \\ \scriptstyle{-2\sqrt{2}} &\scriptstyle{-\frac{4}{3}}\end{array} \Big{)}$ &  $\Big{(}\begin{array}{cc}\scriptstyle{\frac{2}{3}} & \scriptstyle{0} \\ \scriptstyle{0} &\scriptstyle{-\frac{16}{3} }\end{array} \Big{)}$    \\
\cline{1-3} $\begin{array}{c}\scriptstyle{\lambda^c_1\lambda^c_4} \\ \scriptstyle{=\lambda^c_2\lambda^c_3} \end{array} $ &  $\Big{(}\begin{array}{cc}\scriptstyle{-\frac{10}{3}} & \scriptstyle{2\sqrt{2}} \\ \scriptstyle{2\sqrt{2}} &\scriptstyle{-\frac{4}{3}}\end{array} \Big{)}$ &  $\Big{(}\begin{array}{cc}\scriptstyle{-\frac{14}{3}} & \scriptstyle{-\frac{4\sqrt{2}}{3}} \\ \scriptstyle{-\frac{4\sqrt{2}}{3}} &\scriptstyle{0}\end{array} \Big{)}$    \\
\hline\hline
\end{tabular}
\end{table}

\section{Spin states for tetraquark}
In this section, we investigate spin states for tetraquark to calculate the matrix element of the Hamiltonian in Eq.~(\ref{Hamiltonian}).
The case for the spin operators can be treated similarly  as before in that $SU(2)$ is a subgroup of $SU(3)$. A point that is different from the case of $SU(3)$ is that the $SU(2)$ has only one Casimir operator. The only Casimir operator, ${\sigma}\cdot{\sigma}$ classifies the multiplets of $SU(2)$ by the eigenvalues. We describe the explicit form of the total spin S=0,and 1.
The two orthonormal basis states $(\chi_{12})_{s=1}\otimes(\chi_{34})_{s=1}$ and $(\chi_{12})_{s=0}\otimes(\chi_{34})_{s=0}$ with the total S=0 in Eq.~(\ref{basis-S0}) can be expressed as,
 \begin{align}
&(\chi_{12})_{s=1}\otimes(\chi_{34})_{s=1}\nonumber\\
&=\frac{1}{\sqrt{3}}\uparrow(1)\uparrow(2)\otimes\downarrow(3)\downarrow(4)\nonumber\\
&+\frac{1}{\sqrt{3}}\downarrow(1)\downarrow(2)\otimes\uparrow(3)\uparrow(4)\nonumber\\
&-\frac{1}{\sqrt{3}}\frac{1}{\sqrt{2}}(\uparrow(1)\downarrow(2)+\downarrow(1)\uparrow(2))\otimes
\frac{1}{\sqrt{2}}(\uparrow(3)\downarrow(4)+\nonumber\\
&\downarrow(3)\uparrow(4))\nonumber\\
&=\frac{1}{\sqrt{12}}(2\uparrow\uparrow\downarrow\downarrow+2\downarrow\downarrow\uparrow\uparrow
-\uparrow\downarrow
\uparrow\downarrow-\uparrow\downarrow\downarrow\uparrow-\downarrow\uparrow
\uparrow\downarrow-\downarrow\uparrow\downarrow\uparrow)\nonumber,
\end{align}
\begin{align}
&(\chi_{12})_{s=0}\otimes(\chi_{34})_{s=0}\nonumber\\
&=\frac{1}{\sqrt{2}}(\uparrow(1)\downarrow(2)-\downarrow(1)\uparrow(2))\otimes
\frac{1}{\sqrt{2}}(\uparrow(3)\downarrow(4)-\nonumber\\
&\downarrow(3)\uparrow(4))\nonumber\\
&=\frac{1}{2}(\uparrow\downarrow\uparrow\downarrow-
\uparrow\downarrow\downarrow\uparrow-
\downarrow\uparrow\uparrow\downarrow+\downarrow\uparrow\downarrow\uparrow).
\label{s1}
\end{align}
Here, we define spinors as,\\
\begin{equation}
\left(\begin{array}{c}1\\0\end{array} \right)=\uparrow,\qquad
\left(\begin{array}{c}0\\1\end{array} \right)=\downarrow.
\end{equation}
The coefficients appearing in Eq.~(\ref{s1}) are obtained from the Clebsch-Gordan coefficients of $SU(2)$. The three basis states with the total S=1 in Eq.~(\ref{basis-S1}) are given by,
\begin{align}
(\chi_{12})_{s=1}\otimes&(\chi_{34})_{s=0}=\nonumber\\
&\uparrow(1)\uparrow(2)\otimes\frac{1}{\sqrt{2}}
(\uparrow(3)\downarrow(4)-\downarrow(3)\uparrow(4))\nonumber\\
&=\frac{1}{\sqrt{2}}(\uparrow\uparrow\uparrow\downarrow
-\uparrow\uparrow\downarrow\uparrow)\nonumber,
\end{align}
\begin{align}
(\chi_{12})_{s=1}&\otimes(\chi_{34})_{s=1}=\nonumber\\
&\frac{1}{\sqrt{2}}\frac{1}{\sqrt{2}}((\uparrow(1)\downarrow(2)+\downarrow(1)\uparrow(2))\otimes
\uparrow(3)\uparrow(4)\nonumber\\
&-\frac{1}{\sqrt{2}}\uparrow(1)\uparrow(2)\otimes\frac{1}{\sqrt{2}}((\uparrow(3)\downarrow(4)+
\downarrow(3)\uparrow(4))\nonumber\\
&=\frac{1}{2}(\uparrow\downarrow\uparrow\uparrow+\downarrow\uparrow\uparrow\uparrow-
\uparrow\uparrow\uparrow\downarrow-\uparrow\uparrow\downarrow\uparrow)
\nonumber,
\end{align}
\begin{align}
(\chi_{12})_{s=0}\otimes&(\chi_{34})_{s=1}=\nonumber\\
&\frac{1}{\sqrt{2}}(\uparrow(1)\downarrow(2)-\downarrow(1)\uparrow(2))\otimes\uparrow(3)\uparrow(4)
\nonumber\\
&=\frac{1}{\sqrt{2}}(\uparrow\downarrow\uparrow\uparrow
-\downarrow\uparrow\uparrow\uparrow).
\end{align}

It is easy to obtain the result of applying the spin operator ${\sigma}_i\cdot{\sigma}_j$ on these bases through the well known eigenvalues of the Casimir. By analogy with the case of $SU(3)$,
we can find the matrix of ${\sigma}_i\cdot{\sigma}_j$ for S=0 and S=1.
\begin{table}
\caption{The matrix of ${\sigma}_i\cdot{\sigma}_j$ is written in terms of two basis set, $\phi_1=(\chi_{12})_{s=1}\otimes(\chi_{34})_{s=1}
$ and $\phi_2=(\chi_{12})_{s=0}\otimes(\chi_{34})_{s=0}$ for the total S=0, $\psi_1=(\chi_{12})_{s=1}\otimes(\chi_{34})_{s=0}$, $\psi_2= (\chi_{12})_{s=1}\otimes(\chi_{34})_{s=1}$ and $\psi_3=(\chi_{12})_{s=0}\otimes(\chi_{34})_{s=1}$ for the total S=1.}
\begin{tabular}{|c|c|c|}
\hline\hline
& spin=0  states& spin=1 states\\
\cline{1-3}  ${\sigma}_1\cdot{\sigma}_2$
&  $\Big{(}\begin{array}{cc}\scriptstyle{1} & \scriptstyle{0} \\ \scriptstyle{0} &\scriptstyle{-3}\end{array} \Big{)}$ &  $\bigg{(}\begin{array}{ccc}\scriptstyle{1} & \scriptstyle{0}&\scriptstyle{0} \\ \scriptstyle{0}& \scriptstyle{1}&\scriptstyle{0}\\ \scriptstyle{0}&\scriptstyle{0 }&\scriptstyle{-3}\end{array} \bigg{)}$    \\
\cline{1-3} ${\sigma}_1\cdot{\sigma}_3$ &  $\Big{(}\begin{array}{cc}\scriptstyle{-2} & \scriptstyle{-\sqrt{3}} \\ \scriptstyle{-\sqrt{3}} &\scriptstyle{0}\end{array} \Big{)}$ & $\bigg{(}\begin{array}{ccc}\scriptstyle{0} & \scriptstyle{-\sqrt{2}}&\scriptstyle{1} \\ \scriptstyle{-\sqrt{2}}& \scriptstyle{-1}&\scriptstyle{\sqrt{2}}\\ \scriptstyle{1}&\scriptstyle{\sqrt{2}}&\scriptstyle{0}\end{array} \bigg{)}$      \\
\cline{1-3} ${\sigma}_1\cdot{\sigma}_4$ &  $\Big{(}\begin{array}{cc}\scriptstyle{-2} & \scriptstyle{\sqrt{3}} \\ \scriptstyle{\sqrt{3}} &\scriptstyle{0}\end{array} \Big{)}$ &$\bigg{(}\begin{array}{ccc}\scriptstyle{0} & \scriptstyle{\sqrt{2}}&\scriptstyle{-1} \\ \scriptstyle{\sqrt{2}}& \scriptstyle{-1}&\scriptstyle{\sqrt{2}}\\ \scriptstyle{-1}&\scriptstyle{\sqrt{2}}&\scriptstyle{0}\end{array} \bigg{)}$     \\
\cline{1-3} ${\sigma}_2\cdot{\sigma}_3$ &  $\Big{(}\begin{array}{cc}\scriptstyle{-2} & \scriptstyle{\sqrt{3}} \\ \scriptstyle{\sqrt{3}} &\scriptstyle{0}\end{array} \Big{)}$ &  $\bigg{(}\begin{array}{ccc}\scriptstyle{0} & \scriptstyle{-\sqrt{2}}&\scriptstyle{-1} \\ \scriptstyle{-\sqrt{2}}& \scriptstyle{-1}&\scriptstyle{-\sqrt{2}}\\ \scriptstyle{-1}&\scriptstyle{-\sqrt{2}}&\scriptstyle{0}\end{array} \bigg{)}$    \\
\cline{1-3} ${\sigma}_2\cdot{\sigma}_4$ &  $\Big{(}\begin{array}{cc}\scriptstyle{-2} & \scriptstyle{-\sqrt{3}} \\ \scriptstyle{-\sqrt{3}} &\scriptstyle{0}\end{array} \Big{)}$ & $\bigg{(}\begin{array}{ccc}\scriptstyle{0} & \scriptstyle{\sqrt{2}}&\scriptstyle{1} \\ \scriptstyle{\sqrt{2}}& \scriptstyle{-1}&\scriptstyle{-\sqrt{2}}\\ \scriptstyle{1}&\scriptstyle{-\sqrt{2}}&\scriptstyle{0}\end{array} \bigg{)}$   \\
\cline{1-3} ${\sigma}_3\cdot{\sigma}_4$ &  $\Big{(}\begin{array}{cc}\scriptstyle{1} & \scriptstyle{0} \\ \scriptstyle{0} &\scriptstyle{-3}\end{array} \Big{)}$ & $\bigg{(}\begin{array}{ccc}\scriptstyle{-3} & \scriptstyle{0}&\scriptstyle{0} \\ \scriptstyle{0}& \scriptstyle{1}&\scriptstyle{0}\\ \scriptstyle{0}&\scriptstyle{0}&\scriptstyle{1}\end{array} \bigg{)}$    \\
\hline\hline
\end{tabular}
\end{table}
It follows immediately that we can find the matrix of interaction Hamiltonian in Eq.~(\ref{interaction}) for scalar tetraquark and  axial tetraquark. They are obtained from the Kronecker product of the matrix of the color operator $\lambda^c_i\lambda^c_j$ and the spin operator ${\sigma}_i\cdot{\sigma}_j$.
The basis set of the Kronecker product of the matrix of the color operator $\lambda^c_i\lambda^c_j$ and the spin operator ${\sigma}_i\cdot{\sigma}_j$ for scalar tetraquark is given by,
\begin{align}
&\phi_1=(q_1q_2)^{6}_{1}\otimes(\bar{q}_3\bar{q}_4)^{\bar{6}}_{1},
\phi_2=(q_1q_2)^{6}_{0}\otimes(\bar{q}_3\bar{q}_4)^{\bar{6}}_{0},\nonumber\\
&\phi_3=(q_1q_2)^{\bar{3}}_{1}\otimes(\bar{q}_3\bar{q}_4)^{3}_{1},
\phi_4=(q_1q_2)^{\bar{3}}_{0}\otimes(\bar{q}_3\bar{q}_4)^{3}_{0}.\label{scalar-1}
\end{align}
and, the basis set for axial tetraquark is,
\begin{align}
&\phi_1=(q_1q_2)^{6}_{1}\otimes(\bar{q}_3\bar{q}_4)^{\bar{6}}_{0},
\phi_2=(q_1q_2)^{6}_{1}\otimes(\bar{q}_3\bar{q}_4)^{\bar{6}}_{1},\nonumber\\
&\phi_3=(q_1q_2)^{6}_{0}\otimes(\bar{q}_3\bar{q}_4)^{\bar{6}}_{1},
\phi_4=(q_1q_2)^{\bar{3}}_{1}\otimes(\bar{q}_3\bar{q}_4)^{3}_{0},\nonumber\\
&\phi_5=(q_1q_2)^{\bar{3}}_{1}\otimes(\bar{q}_3\bar{q}_4)^{3}_{1},
\phi_6=(q_1q_2)^{\bar{3}}_{0}\otimes(\bar{q}_3\bar{q}_4)^{3}_{1}.
\end{align}
The matrix of interaction Hamiltonian in Eq.~(\ref{interaction}) for scalar tetraquark in terms of the color-spin basis states is written as,
\begin{align}
&\langle-\sum_{i<j}C_{ij}\lambda^c_i\lambda^c_j{\sigma}_i\cdot{\sigma}_j\rangle
=H^{\prime}_{CM}=\nonumber\\
&-C_{12}\left(\begin{array}{cc}\scriptstyle{\frac{4}{3}}\Big{(}\begin{array}{cc}\scriptstyle{1} & \scriptstyle{0} \\ \scriptstyle{0} &\scriptstyle{-3}\end{array} \Big{)} & \scriptstyle{0}\Big{(}\begin{array}{cc}\scriptstyle{1} & \scriptstyle{0} \\ \scriptstyle{0} &\scriptstyle{-3}\end{array} \Big{)} \\ \scriptstyle{0}\Big{(}\begin{array}{cc}\scriptstyle{1} & \scriptstyle{0} \\ \scriptstyle{0} &\scriptstyle{-3}\end{array} \Big{)} &\scriptstyle{-\frac{8}{3}}\Big{(}\begin{array}{cc}\scriptstyle{1} & \scriptstyle{0} \\ \scriptstyle{0} &\scriptstyle{-3}\end{array} \Big{)}\end{array} \right)+\nonumber\\
&-C_{13}\left(\begin{array}{cc}\scriptstyle{-\frac{10}{3}}\Big{(}\begin{array}{cc}\scriptstyle{-2} & \scriptstyle{-\sqrt{3}} \\ \scriptstyle{-\sqrt{3}} &\scriptstyle{0}\end{array} \Big{)} & \scriptstyle{-2\sqrt{2}}\Big{(}\begin{array}{cc}\scriptstyle{-2} & \scriptstyle{-\sqrt{3}} \\ \scriptstyle{-\sqrt{3}} &\scriptstyle{0}\end{array} \Big{)} \\ \scriptstyle{-2\sqrt{2}}\Big{(}\begin{array}{cc}\scriptstyle{-2} & \scriptstyle{-\sqrt{3}} \\ \scriptstyle{-\sqrt{3}} &\scriptstyle{0}\end{array} \Big{)} &\scriptstyle{-\frac{4}{3}}\Big{(}\begin{array}{cc}\scriptstyle{-2} & \scriptstyle{-\sqrt{3}} \\ \scriptstyle{-\sqrt{3}} &\scriptstyle{0}\end{array} \Big{)}\end{array} \right)+\nonumber
\end{align}
\begin{align}
&-C_{14}\left(\begin{array}{cc}\scriptstyle{-\frac{10}{3}}\Big{(}\begin{array}{cc}\scriptstyle{-2} & \scriptstyle{\sqrt{3}} \\ \scriptstyle{\sqrt{3}} &\scriptstyle{0}\end{array} \Big{)} & \scriptstyle{2\sqrt{2}}\Big{(}\begin{array}{cc}\scriptstyle{-2} & \scriptstyle{\sqrt{3}} \\ \scriptstyle{\sqrt{3}} &\scriptstyle{0}\end{array} \Big{)} \\ \scriptstyle{2\sqrt{2}}\Big{(}\begin{array}{cc}\scriptstyle{-2} & \scriptstyle{\sqrt{3}} \\ \scriptstyle{\sqrt{3}} &\scriptstyle{0}\end{array} \Big{)} &\scriptstyle{-\frac{4}{3}}\Big{(}\begin{array}{cc}\scriptstyle{-2} & \scriptstyle{\sqrt{3}} \\ \scriptstyle{\sqrt{3}} &\scriptstyle{0}\end{array} \Big{)}\end{array} \right)+\nonumber
\end{align}
\begin{align}
&-C_{23}\left(\begin{array}{cc}\scriptstyle{-\frac{10}{3}}\Big{(}\begin{array}{cc}\scriptstyle{-2} & \scriptstyle{\sqrt{3}} \\ \scriptstyle{\sqrt{3}} &\scriptstyle{0}\end{array} \Big{)} & \scriptstyle{2\sqrt{2}}\Big{(}\begin{array}{cc}\scriptstyle{-2} & \scriptstyle{\sqrt{3}} \\ \scriptstyle{\sqrt{3}} &\scriptstyle{0}\end{array} \Big{)} \\ \scriptstyle{2\sqrt{2}}\Big{(}\begin{array}{cc}\scriptstyle{-2} & \scriptstyle{\sqrt{3}} \\ \scriptstyle{\sqrt{3}} &\scriptstyle{0}\end{array} \Big{)} &\scriptstyle{-\frac{4}{3}}\Big{(}\begin{array}{cc}\scriptstyle{-2} & \scriptstyle{\sqrt{3}} \\ \scriptstyle{\sqrt{3}} &\scriptstyle{0}\end{array} \Big{)}\end{array} \right)+\nonumber
\end{align}
\begin{align}
&-C_{24}\left(\begin{array}{cc}\scriptstyle{-\frac{10}{3}}\Big{(}\begin{array}{cc}\scriptstyle{-2} & \scriptstyle{-\sqrt{3}} \\ \scriptstyle{-\sqrt{3}} &\scriptstyle{0}\end{array} \Big{)} & \scriptstyle{-2\sqrt{2}}\Big{(}\begin{array}{cc}\scriptstyle{-2} & \scriptstyle{-\sqrt{3}} \\ \scriptstyle{-\sqrt{3}} &\scriptstyle{0}\end{array} \Big{)} \\ \scriptstyle{-2\sqrt{2}}\Big{(}\begin{array}{cc}\scriptstyle{-2} & \scriptstyle{-\sqrt{3}} \\ \scriptstyle{-\sqrt{3}} &\scriptstyle{0}\end{array} \Big{)} &\scriptstyle{-\frac{4}{3}}\Big{(}\begin{array}{cc}\scriptstyle{-2} & \scriptstyle{-\sqrt{3}} \\ \scriptstyle{-\sqrt{3}} &\scriptstyle{0}\end{array} \Big{)}\end{array} \right)+\nonumber
\end{align}
\begin{align}
&-C_{34}\left(\begin{array}{cc}\scriptstyle{\frac{4}{3}}\Big{(}\begin{array}{cc}\scriptstyle{1} & \scriptstyle{0} \\ \scriptstyle{0} &\scriptstyle{-3}\end{array} \Big{)} & \scriptstyle{0}\Big{(}\begin{array}{cc}\scriptstyle{1} & \scriptstyle{0} \\ \scriptstyle{0} &\scriptstyle{-3}\end{array} \Big{)} \\ \scriptstyle{0}\Big{(}\begin{array}{cc}\scriptstyle{1} & \scriptstyle{0} \\ \scriptstyle{0} &\scriptstyle{-3}\end{array} \Big{)} &\scriptstyle{-\frac{8}{3}}\Big{(}\begin{array}{cc}\scriptstyle{1} & \scriptstyle{0} \\ \scriptstyle{0} &\scriptstyle{-3}\end{array} \Big{)}\end{array} \right).
\end{align}
To compare with the result which can be found in Ref~\cite{Buccella:2006fn}, we change the basis set in Eq.~(\ref{scalar-1}) into,
\begin{align}
&\phi_1=(q_1q_2)^{6}_{1}\otimes(\bar{q}_3\bar{q}_4)^{\bar{6}}_{1},
\phi_2=(q_1q_2)^{\bar{3}}_{0}\otimes(\bar{q}_3\bar{q}_4)^{3}_{0},\nonumber\\
&\phi_3=(q_1q_2)^{6}_{0}\otimes(\bar{q}_3\bar{q}_4)^{\bar{6}}_{0},
\phi_4=(q_1q_2)^{\bar{3}}_{1}\otimes(\bar{q}_3\bar{q}_4)^{3}_{1},\label{scalar-2}
\end{align}
Then, it is found that the transformation from the basis set in Eq.~(\ref{scalar-1}) to the basis set in Eq.~(\ref{scalar-2}) is,
\begin{equation}
U=\left(\begin{array}{cccc}1 &0&0&0\\
0&0&1&0\\
0&0&0&1\\
0&1&0&0\end{array} \right).
\end{equation}
The matrix of the interaction Hamiltonian in Eq.~(\ref{interaction}) denoted by $H_{CM}$ for scalar tetraqurak is acquired by the similarity transformation :
\begin{equation}
H_{CM}=U^TH^{\prime}_{CM}U=
-\left(\begin{array}{cc}A & B \\
C &D\end{array} \right),
\end{equation}
with 2 by 2 submatrices
\begin{align}
&A_{11}=\frac{4}{3}(C_{12}+C_{34})+\frac{20}{3}(C_{13}+C_{14}+C_{23}+C_{24}),\nonumber\\
&A_{12}=A_{21}=2\sqrt{6}(C_{13}+C_{14}+C_{23}+C_{24}),\nonumber\\
&A_{22}=8(C_{12}+C_{34})\nonumber,
\end{align}
\begin{equation}
B=C^T=\frac{2}{\sqrt{3}}(C_{13}-C_{14}-C_{23}+C_{24})\Big{(}\begin{array}{cc}\scriptstyle{5} & \scriptstyle{2\sqrt{6}} \\ \scriptstyle{0} &\scriptstyle{2}\end{array} \Big{)}\nonumber,
\end{equation}
\begin{align}
&D_{11}=-4(C_{12}+C_{34}),\nonumber\\
&D_{12}=D_{21}=2\sqrt{6}(C_{13}+C_{14}+C_{23}+C_{24}),\nonumber\\
&D_{22}=-\frac{8}{3}(C_{12}+C_{34}-C_{13}-C_{14}-C_{23}-C_{24}).
\end{align}
In a situation where $C_{13}=C_{23}$ and $C_{14}=C_{24}$, the matrix of interaction Hamiltonian in Eq.~(\ref{interaction}) for scalar tetraquarks reduces to the block diagonal form,
\begin{equation}
H_{CM}=
-\left(\begin{array}{cc}A & 0 \\
0&D\end{array} \right).
\end{equation}
This means that the flavor-symmetry of light diquark causes the separation of $\bar{3}_f$ and $6_f$. We can apply  the same procedure to calculate the matrix of the interaction Hamiltonian in Eq.~(\ref{interaction}) for axial tetraquark.

\textit{Acknowledgements}
We would like to thank S. Takeuchi for useful discussions.  This work was supported by  the Korean Research Foundation under Grant Nos. KRF-2011-0020333  and KRF-2011-0030621.

\end{document}